\documentclass[prd,superscriptaddress,altaffilletter,nofootinbib]{revtex4}

\usepackage{cancel}
\usepackage{graphicx}
\usepackage{amsmath}
\usepackage{amssymb}
\usepackage{graphicx,epsfig}
\usepackage[dvipsnames]{xcolor}
\newcommand{\be}{\begin{equation}}
\newcommand{\ee}{\end{equation}}
\newcommand{\bea}{\begin{eqnarray}}
\newcommand{\eea}{\end{eqnarray}}
\newcommand{\der}{\partial}
\newcommand{\vphi}{\varphi}

\begin{document}

%%%%%%%%%%%%%%%%%%%%%%%%%%

%%%%%%%%%%%%%%%%%%%%%%%%%%

\title{Conformal form-invariant parametrization of scalar-tensor gravity theories: A critical analysis}

\author{Israel Quiros}\email{iquiros@fisica.ugto.mx;i.quiros@ugto.mx}\affiliation{Departamento Ingenier\'ia Civil, Divisi\'on de Ingenier\'ia, Universidad de Guanajuato, C.P. 36000, Gto., M\'exico.}

\author{Amit Kumar Rao}\email{amit.kumar@ugto.mx;  amit.akrao@gmail.com}\affiliation{Departamento Ingenier\'ia Civil, Divisi\'on de Ingenier\'ia, Universidad de Guanajuato, C.P. 36000, Gto., M\'exico.}\affiliation{Instituto de F\'{\i}sica, Benem\'erita Universidad Aut\'onoma de Puebla,\\
Apartado Postal J-48, 72570, Puebla, Puebla, M\'exico.}

\date{\today}

\begin{abstract} Based on the recent result that, if the masses of timelike fields are point-dependent fields themselves, the action of matter fields is conformal form-invariant in its standard form, and on the active and passive approaches to conformal transformations, we review the conformal form-invariant parametrization of scalar-tensor gravity theories. We investigate whether this parametrization is actually different from other existing parametrizations. We also check the universality of the claim that the classical physical predictions of these theories are conformal-frame invariants.\end{abstract}

%\pacs{04.50.Kd, 04.50.Cd, 11.10.Ef, 98.80.-k, 98.80.Jk}

%%%%%%%%%%%%%%%%%%%%%%%%%%%%%

\maketitle

%%%%%%%%%%%%%%%%%%%%%%%%%%%%%

%------Beginning of draft----------

%%%%%%%%%%%%%%%%%%%%%%%%%%%%%%%%%%%

\section{Introduction}
\label{sect-intro}

%%%%%%%%%%%%%%%%%%%%%%%%%%%%%%%%%%

The conformal frames issue (CFI) arises from different, sometimes opposite, understandings of the conformal transformation (CT) of the metric \cite{dicke-1962, morganstern_1970, anderson-1971, bekenstein_1980, bekenstein_1993, cotsakis_1993, magnano_1994, capozziello_1997, kaloper_prd_1998, faraoni_rev, faraoni_ijmpd_1999, quiros_prd_2000, fabris_2000, casadio_2002, alvarez_2002, vollick_cqg_2004, fujii_book, faraoni_book, flanagan_cqg_2004, bhadra_2007, hassaine_2007, fujii_2007, faraoni_prd_2007, catena_prd_2007, sotiriou_ijmpd_2008, odi-1, elizalde_grg_2010, deruelle_2011, chiba_2013, capozziello_prd_2013, quiros_grg_2013, jarv_2015, karamitsos_2016, sasaki_2016, banerjee_2016, pandey_2017, karamitsos_2018, quiros_ijmpd_2019, quiros_ijmpd_2020, gionti_2021, pal_2022, shtanov-2022, bamber_prd_2023, moinuddin_epjc_2023, mukherjee_epjc_2023, mazumdar_prd_2023, bamba_prd_2024, gionti_epjc_2024, mandal_2024, quiros-2025-a, quiros-2025-b} when applied within the context of scalar-tensor gravitational (STG) theories \cite{brans-dicke-1961, bergmann_1968, nordvedt_1970, wagoner_1970, ross_1972, fujii-1974, isenberg_1976, casas_1992, fujii_book, faraoni_book, quiros_ijmpd_2019}.\footnote{In this work, we consider ``traditional'' STG theories. We leave Horndeski theories for future work.} Mathematically, CT can be stated as the following transformation of the metric tensor:

\begin{align} g_{\mu\nu}\rightarrow&\hat g_{\mu\nu}=\Omega^2g_{\mu\nu}\;\left(g^{\mu\nu}\rightarrow\hat g^{\mu\nu}=\Omega^{-2}g^{\mu\nu}\right),\label{conf-t}\end{align} where the positive smooth function $\Omega=\Omega(x)$ is the conformal factor, simultaneously with suitable transformations of the other fields. In fact, CT acts only on fields, so it does not affect either the spacetime coordinates or the spacetime points; that is, this is not a spacetime diffeomorphism.

%-----------------------------------

STG theories can be formulated in terms of different sets of fields called ``frames'', which are related by CT \eqref{conf-t}, simultaneously with an appropriate transformation of the remaining fields present. Among these, the Einstein frame (EF) and the Jordan frame (JF) play an important role. There is a fairly well-accepted hypothesis that physical laws must be invariant under conformal transformations \cite{faraoni_prd_2007, catena_prd_2007} or, in equivalent words, that physics must be invariant under local transformations of units \cite{dicke-1962}. However, there has been a long-standing confusion about the physical equivalence of the different conformal frames, known as the CFI \cite{dicke-1962, morganstern_1970, anderson-1971, bekenstein_1980, bekenstein_1993, cotsakis_1993, magnano_1994, capozziello_1997, kaloper_prd_1998, faraoni_rev, faraoni_ijmpd_1999, quiros_prd_2000, fabris_2000, casadio_2002, alvarez_2002, vollick_cqg_2004, fujii_book, faraoni_book, flanagan_cqg_2004, bhadra_2007, hassaine_2007, fujii_2007, faraoni_prd_2007, catena_prd_2007, sotiriou_ijmpd_2008, odi-1, elizalde_grg_2010, deruelle_2011, chiba_2013, capozziello_prd_2013, quiros_grg_2013, jarv_2015, karamitsos_2016, sasaki_2016, banerjee_2016, pandey_2017, karamitsos_2018, quiros_ijmpd_2019, quiros_ijmpd_2020, gionti_2021, pal_2022, shtanov-2022, bamber_prd_2023, moinuddin_epjc_2023, mukherjee_epjc_2023, mazumdar_prd_2023, bamba_prd_2024, gionti_epjc_2024, mandal_2024, quiros-2025-a, quiros-2025-b}.

An interesting idea for discussing the above issue has been presented in \cite{flanagan_cqg_2004}. In this bibliographic reference, a formalism has been proposed, which is based on a conformal form-invariant parametrization of STG theory. According to the proposed formalism, the CFI does not arise, since the different conformal frames (JF, EF, etc.) do not occur.\footnote{Other authors have further developed this formalism by looking for conformal invariant measured quantities \cite{catena_prd_2007, jarv_2015}.} If we adopt the parametrization proposed in \cite{flanagan_cqg_2004}, the action and the derived equations of motion (EOM) are form-invariant under an extended group of conformal transformations that includes transformations of the field-dependent parameters (see Section \ref{sect-flan}). In \cite{quiros-2025-b} it has been concluded that the CFI arises if the STG theory in the JF Brans-Dicke (BD) parametrization is submitted to a restricted group of conformal transformations that does not include the transformation of the coupling parameter $\omega=\omega(\phi)$. If the latter transformation is included in the conformal transformation of fields and field-dependent parameters, the STG theory in JFBD parametrization is conformal form-invariant so that the different conformal frames do not arise. It is further argued that, since the transformation of the BD field $\phi\rightarrow\Omega^{-2}\phi$ is included in the CT group, any parameter dependent on $\phi$, such as the coupling function $\omega=\omega(\phi)$, must also be transformed. Then the arising of the conformal frames issue is a consequence of incorrectly omitting the transformation of the coupling parameter. 

The above conclusion is correct, however, only in vacuum \cite{quiros-2025-b}. When timelike matter fields are coupled to gravity, it is required, in addition, that the masses of fields transform themselves as fields under CTs \cite{dicke-1962}: 

\begin{align} m\to\hat m=\Omega^{-1}m.\label{mass-t}\end{align} Based on the dimensional analysis, it can be shown that the transformation of masses according to \eqref{mass-t} leads to the energy density of perfect fluids transforming as,

\begin{align} \rho\to\hat\rho=\Omega^{-4}\rho,\label{rho-t}\end{align} under CTs. In general we have that $m=m(\phi)$, and the energy density of fluids $\rho=\rho(\phi)$, are functions of the BD field, so the Ward identity $\delta{\cal L}_m/\delta\phi=\sqrt{-g}\,T^{(m)}/2\phi$, is required to hold, where ${\cal L}_m$ is the Lagrangian density of matter and $T^{(m)}=g^{\mu\nu}T^{(m)}_{\mu\nu}$ is the trace of the stress-energy tensor (SET) of matter. Fulfillment of the Ward identity is a cornerstone to get to the correct equations of motion (EOM) \cite{quiros-2025-b}.

Although the conformal form-invariant parametrization proposed in \cite{flanagan_cqg_2004} and further developed in \cite{catena_prd_2007, jarv_2015} seems to show that all classical physical predictions of STG theories are conformal-frame invariants, this conclusion reflects only one aspect of the conformal transformation, which we call the passive approach to conformal transformations (PACT). It has been shown \cite{quiros-2025-a, quiros-2025-b} that the active approach to conformal transformations (AACT) must also be considered. The separation of CTs into passive and active transformations can be illustrated if conformal transformations of fields are regarded as transformations of ``generalized coordinates'' in the configuration space. We can trace a parallel with the geometric approach to multifield inflation proposed in \cite{gong_2011}, where the scalar fields $\vphi_a$ ($a=1,2,...,N$) are treated as coordinates living in some field-space manifold, so that any transformation of the scalar fields is regarded as a coordinate transformation in the field space (see also \cite{jarv_2015, karamitsos_2016, karamitsos_2018, karamitsos_2020, karamitsos_2021}). This approach is further developed in references \cite{quiros-2025-a, quiros-2025-b} by assuming that not only the scalar field $\phi$, but also the metric $g_{\mu\nu}$ and the matter fields $\chi$, are generalized coordinates in the configuration space manifold ${\cal M}_\text{fields}$, where each point represents a global gravitational state (GGS) of the system.\footnote{By global gravitational state we understand complete knowledge of the metric $g_{\mu\nu}$ and of the scalar field $\phi$, as well as of any matter fields $\chi$, at any spacetime point. In what follows, we refer to this indistinctly as the global gravitational state or just the gravitational state.} Consider the following CTs of the fields:

\begin{align} g_{\mu\nu}\rightarrow g'_{\mu\nu}=\Omega^2g_{\mu\nu},\;\phi\rightarrow\phi'=\Omega^{-2}\phi,\;\chi\rightarrow\chi'=\Omega^{w_\chi}\chi,\label{gauge-t}\end{align} where $\chi$ is the collective name for a set of $N$ matter fields $\chi=\{\chi_1,\chi_2,...,\chi_N\}$ and $w_\chi$ is its conformal weight. To understand the differences between the PACT and the AACT, it is essential to distinguish between different representations of the same gravitational state and different gravitational states in ${\cal M}_\text{fields}$. 

%---------passive---------------

Following the PACT, let ${\cal R}_g:(g_{\mu\nu},\phi,\chi)$ and ${\cal R}'_g:(g'_{\mu\nu},\phi',\chi'),$ be different representations of the same gravitational state ${\cal S}_\mathfrak{g}:(\Psi,\mathfrak{g}_{\mu\nu})$ in ${\cal M}_\text{fields}$, where the conformal invariant composite quantities,

\begin{align} \mathfrak{g}_{\mu\nu}=\frac{\phi}{M^2_\text{pl}}\,g_{\mu\nu},\;\Psi=\left(\frac{\phi}{M^2_\text{pl}}\right)^\frac{w_\chi}{2}\chi,\label{inv-met}\end{align} represent the physically meaningful metric tensor and matter fields, respectively. Therefore,

\begin{align} \mathfrak{g}_{\mu\nu}=\frac{\phi}{M^2_\text{pl}}\,g_{\mu\nu}=\frac{\phi'}{M^2_\text{pl}}\,g'_{\mu\nu}=\mathfrak{g}'_{\mu\nu},\;\Psi=\left(\frac{\phi}{M^2_\text{pl}}\right)^\frac{w_\chi}{2}\,\chi=\left(\frac{\phi'}{M^2_\text{pl}}\right)^\frac{w_\chi}{2}\,\chi'=\Psi',\label{passive-t}\end{align} so, it is clear that passive CTs \eqref{gauge-t} amount to the identity transformation of the physical fields. Hence, seeking the properties of conformal symmetry under the PACT is not a well-posed problem. This statement will be demonstrated in Section \ref{sect-pact}.

%-----------active----------

In contrast to PACT, active CTs:

\begin{align} g_{\mu\nu}\rightarrow\bar g_{\mu\nu}=\Omega^2g_{\mu\nu},\;\phi\to\bar\phi=\Omega^{-2}\phi,\;\chi\to\bar\chi=\Omega^{w_\chi}\chi,\label{active-t}\end{align} where $\bar\phi=\Omega^{-2}\phi$ $\Rightarrow$ $\Omega^2=\phi/\bar\phi$, relate different GGSs ${\cal S}_g:(g_{\mu\nu},\phi,\chi)$ and $\bar{\cal S}_g:(\bar g_{\mu\nu},\bar\phi,\bar\chi)$ in ${\cal M}_\text{fields}$, or, since $\phi$ is a gauge field, we can say that the ``old'' and ``new'' gravitational states: ${\cal S}_g:(g_{\mu\nu},\chi)$ and $\bar{\cal S}_g:(\bar g_{\mu\nu},\bar\chi)$, respectively, are linked by the following relationships:

\begin{align} \bar g_{\mu\nu}=\frac{\phi}{\bar\phi}\,g_{\mu\nu},\;\bar\chi=\left(\frac{\phi}{\bar\phi}\right)^\frac{w_\chi}{2}\chi,\label{active-t'}\end{align} which follow from the invariants $(\phi/M^2_\text{pl}) g_{\mu\nu}=(\bar\phi/M^2_\text{pl}) \bar g_{\mu\nu}=$ inv., and $(\phi/M^2_\text{pl})^{w_\chi/2}\chi=(\bar\phi/M^2_\text{pl})^{w_\chi/2}\bar\chi=$ inv. The fields $g_{\mu\nu}$ and $\chi$ are physical, contrary to the PACT, where the physical fields are the conformal invariant quantities $\mathfrak{g}_{\mu\nu}$ and $\Psi$. In the latter case, the fields $g_{\mu\nu}$, $\phi$, and $\chi$ are auxiliary fields without immediate physical meaning.

%------------------

\begin{figure*}[t!]\centering
\includegraphics[width=8.5cm]{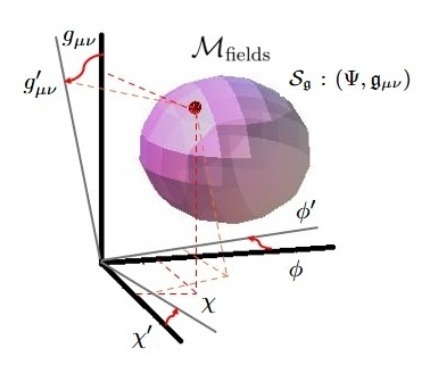}
\includegraphics[width=8.5cm]{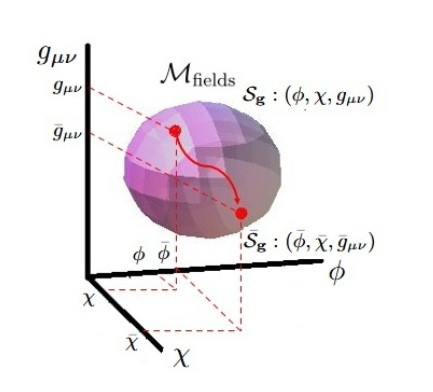}
%\vspace{1.2cm}
\caption{Drawings of the configuration space ${\cal M}_\text{fields}$ (shadowed solid region in the phase space). Each point in the manifold represents a global gravitational state. In the left figure, the passive approach to CT is illustrated. The conformal transformation amounts to a ``rotation'' of the coordinate system $R:(\phi,\chi,g_{\mu\nu})$ in the phase space, which leaves invariant the gravitational state ${\cal S}_\mathfrak{g}:(\Psi,\mathfrak{g}_{\mu\nu})$, where the invariant metric and mater fields are given by \eqref{inv-met}. The active point of view on the CT is illustrated in the right figure. In this case, the conformal transformation represents an actual ``motion'' of the point $(\phi,\chi,g_{\mu\nu})$ in ${\cal M}_\text{fields}$. That is, it represents a real change of the gravitational state ${\cal S}_{\bf g}:(\phi,\chi,g_{\mu\nu})$ $\rightarrow{\bar{\cal S}}_{\bf g}:(\bar\phi,\bar\chi,\bar g_{\mu\nu})$.}\label{fig1}\end{figure*}

The differentiation between passive and active CT in the configuration space is illustrated in FIG \ref{fig1}. In the left figure, passive CT is viewed as a rotation of the ``coordinate system'' $R:(\chi,\phi,g_{\mu\nu})\rightarrow R':(\chi',\phi',g'_{\mu\nu})$ where the gravitational state ${\cal S}_\mathfrak{g}:(\Psi,\mathfrak{g}_{\mu\nu})$ is unchanged, while in the right figure, active CT is represented as a real ``motion'' in configuration space ${\cal S}_g:(\chi,\phi,g_{\mu\nu})\rightarrow\bar{\cal S}_g:(\bar\chi,\bar\phi,\bar g_{\mu\nu})$, where each point $(\chi,\phi,g_{\mu\nu})$ in ${\cal M}_\text{fields}$ is to be associated with a different GGS.

%------------------------------------------------

%------------------------------

%-------------aim------------------

In this document, we shall demonstrate that, if we adopt the PACT, conformal form-invariance of STG theory in the general parametrization of \cite{flanagan_cqg_2004} (see also \cite{catena_prd_2007, jarv_2015}) is just a spurious or fake symmetry, which is replaced by the identity transformation of the physical fields. This is demonstrated in two steps: First, we write the action of this parameterization in terms of conformal invariant fields $\mathfrak{g}_{\mu\nu}$ and $\Psi$ and, second, we show that the conformal transformations \eqref{passive-t} of the auxiliary fields $g_{\mu\nu},$ $\phi$ and $\chi$ amount to the identity transformation of the conformal invariant fields. In contrast, if we assume the AACT, conformal form-invariance is an actual symmetry of STG theory in the parametrization of \cite{flanagan_cqg_2004}. We conclude that, under PACT, since auxiliary fields are removed in the covariant formulation of the theory on the configuration space, conformal symmetry is fictitious. This means, in turn, that the claim that the predictions of STG theories must be conformal invariants \cite{catena_prd_2007, jarv_2015} is senseless. We shall also show that the quite general parametrization of STG theory proposed in \cite{flanagan_cqg_2004} is fully equivalent to the well-known JFBD parametrization, so no new elements are given in that bibliographic reference, with respect to the earlier work \cite{faraoni_1998}. The generalization of this result to the case when matter fields, including timelike fields, are coupled to gravity is also provided. 

%------------organization---------------

The paper is organized as follows. The fundamentals of the conformal form-invariant parametrization proposed in \cite{flanagan_cqg_2004} are exposed in Section \ref{sect-flan}. In Section \ref{sect-mac}, we criticize the way in which the transformation of the matter action under CTs is discussed in the bibliography. The equivalence of the parametrization proposed in \cite{flanagan_cqg_2004}, for the vacuum, with an earlier proposal \cite{faraoni_1998} based on STG theory in the customary JFBD parametrization, is demonstrated in Section \ref{sect-equiv-vac}, while in Section \ref{sect-equiv-mat} the equivalence is generalized to the case when arbitrary matter fields are coupled to gravity. In Section \ref{sect-math-base}, we discuss the mathematical basis for the passive and active approaches to conformal transformations. The passive approach to the form-invariant parametrization of STG theory is separately discussed in Section \ref{sect-pact}, while in Section \ref{sect-inv-q}, the related subject about invariant quantities in STG theories is discussed. The results obtained in this paper are discussed in Section \ref{sect-discu}, followed by concluding remarks. For completeness of our presentation, we have included an Appendix section that contains some demonstrations which are useful to understand the main text, as well as other relevant material.

%-------------------------------------

%%%%%%%%%%%%%%%%%%%%%%%%%%%%%%%%%%%%%%%%%%

\section{Form-invariant parametrization}
\label{sect-flan}

%%%%%%%%%%%%%%%%%%%%%%%%%%%%%%%%%%%%%%%%%%

In \cite{flanagan_cqg_2004}, a conformal-invariant approach to STG theories was proposed to clarify the conformal frames issue (CFI). In that reference, it is stated that the STG action can be parametrized in the following form:
 
\begin{align} S=\frac{1}{2}\int d^4x\sqrt{-g}\left[AR-B(\der\phi)^2-2V\right]+S_m,\label{flan-lag}\end{align} where we have used the following notation: $(\der\phi)^2\equiv g^{\mu\nu}\der_\mu\phi\der_\nu\phi$, and the action of the matter fields $S_m$ is given by

\begin{align} S_m=S_m[\chi,e^{2\alpha}g_{\mu\nu}]=\int d^4x\sqrt{-g}\,L_m\left(\chi,e^{2\alpha}g_{\mu\nu}\right).\label{flan-mat-action}\end{align} In these expressions, the parameters $A=A(\phi)$, $B=B(\phi)$, the self-interaction potential $V=V(\phi)$, and $\alpha=\alpha(\phi)$, are arbitrary functions of the scalar field $\phi$, while the matter fields are collectively denoted by $\chi=\{\chi_1,\chi_2,...,\chi_N\}$ ($N$ is the total number of matter fields). According to the parametrization \eqref{flan-lag} the relevant conformal frames correspond to particular choices of the parameters $A$, $B$, $V$ and $\alpha$. For example, the JF frame corresponds to the choice $\alpha=0$ and $B=1$. Free-falling objects made of matter fields $\chi$ are said to follow the geodesics of the JF metric. The EF corresponds to the choice $A=1$ and $B=1$. In this case, the free-falling objects follow the geodesics of the Jordan frame, so the occurrence of a physical metric and, correspondingly, of a physical frame makes sense. 

It has been demonstrated in \cite{flanagan_cqg_2004} that the gravitational piece of action \eqref{flan-lag} is form-invariant under CT \eqref{conf-t} and the simultaneous transformations:

\begin{align} &\hat\phi=f(\phi),\;\hat A=\Omega^{-2}A,\;\hat V=\Omega^{-4}V,\;\hat\alpha=\alpha-\ln\Omega,\;\hat B=\Omega^{-2}(f')^{-2}\left[B+6\frac{\Omega'}{\Omega}A'-6\left(\frac{\Omega'}{\Omega}\right)^2A\right],\label{flan-t}\end{align} where we have adopted the conventions of \cite{jarv_2015}, so that a prime in a quantity with a hat denotes derivative with respect to $\hat\phi$, e. g., $\hat{\cal A}'=d\hat{\cal A}/d\hat\phi$, etc., and a prime in a quantity without a hat denotes a derivative with respect to $\phi$, for example: ${\cal A}'=d{\cal A}/d\phi$. In addition, $d\hat\phi=f'd\phi,$ where $f'\equiv\der_\phi f.$ Consequently, the different conformal frames (JF, EF, etc.) do not arise. 

The analysis is a bit more complicated if we consider matter fields. It is seen from \eqref{flan-mat-action} that the matter fields are minimally coupled to the conformal invariant metric 

\begin{align} \mathfrak{g}_{\mu\nu}=e^{2\alpha}g_{\mu\nu}.\label{phys-g}\end{align} This is the metric tensor that defines the proper time measured by clocks. That is, matter fields follow the geodesics of the metric $\mathfrak{g}_{\mu\nu}$ since, as clearly stated in \cite{brans-1988}, the operational significance of the metric must be found in the geometry of spacetime, which is measured by classical particle paths. Besides, as stated in \cite{flanagan_cqg_2004}, the time measured by clocks in the theory \eqref{flan-lag} is the proper time associated with the conformal invariant metric \eqref{phys-g}, i.e., this is the metric with operational meaning or the ``physical metric''. Note that the matter action \eqref{flan-mat-action} is conformal form-invariant only if we replace it by the following: 

\begin{align} S_m=S_m[\Psi,\mathfrak{g}_{\mu\nu}]=\int d^4x\sqrt{-\mathfrak{g}}\,L_m(\Psi,\mathfrak{g}_{\mu\nu}),\nonumber\end{align} where $\Psi\equiv e^{w_\chi\alpha}\chi$, represents the conformal invariant matter fields. This choice of the matter action picks out the passive approach to CTs because only the conformal invariant quantities matter.

The other way in which the coupling of matter to gravity in the parametrization \eqref{flan-lag} is conformal form-invariant is by assuming the transformation \eqref{mass-t} of masses (and \eqref{rho-t} of the energy density) under the AACT. In this case, the standard action of matter fields:

\begin{align} S_m=S_m[\chi,g_{\mu\nu}]=\int d^4x\sqrt{-g}\,L_m\left(\chi,g_{\mu\nu}\right),\nonumber\end{align} is itself conformal form-invariant \cite{quiros-2025-a}. The matter fields $\chi$ are minimally coupled to the metric tensor $g_{\mu\nu}$, so that the latter is the physical metric tensor.

%-------------------------------

%%%%%%%%%%%%%%%%%%%%%%%%%%%%%%%%%%%%%%%%%%%%%%%%%%%%%%%%%%

\section{Criticism of the application of conformal transformation of the matter action in the bibliography}
\label{sect-mac}

%%%%%%%%%%%%%%%%%%%%%%%%%%%%%%%%%%%%%%%%%%%%%%%%%%%%%%%%%%

There is no unanimous treatment of this subject in the bibliography. For example, in \cite{flanagan_cqg_2004}, from equations (2.2) and (2.6) it follows that\footnote{In \cite{flanagan_cqg_2004}, a quite different notation is used. For example, $\psi_m$ is used to account for matter fields instead of $\chi$.}

\begin{align} S_m\left[\chi,e^{2\alpha}g_{\mu\nu}\right]=S_m\left[\chi,e^{2\hat\alpha}\hat g_{\mu\nu}\right],\label{flan-lm-eq}\end{align} while, from equations (1) and (4) of \cite{jarv_2015}, it is seen that\footnote{Notice that there is a subtle difference between \eqref{flan-lm-eq} and \eqref{jarv-lm-eq} in the hat over the matter action.}

\begin{align} S_m\left[\chi,e^{2\alpha}g_{\mu\nu}\right]=\hat S_m\left[\chi,e^{2\hat\alpha}\hat g_{\mu\nu}\right].\label{jarv-lm-eq}\end{align} In both cases, the matter fields considered are not transformed by the conformal transformation, which is correct only for null fields and radiation. In general, the conformal weight of timelike matter fields is non-vanishing. For example, the fermion field $\psi$ has conformal weight $w_\psi=-3/2$. Therefore, either only conformal invariant matter fields (for instance, radiation) couple to gravity in the formalism of \cite{flanagan_cqg_2004}, or the matter action \eqref{flan-lm-eq} is not conformal form-invariant in general.

In other bibliographic references, for example, in \cite{faraoni_rev, bamber_prd_2023, faraoni_book, capozziello_rev_2011}, it is stated that under suitable CTs (the conformal factor chosen to be $\Omega\propto\sqrt\phi$), the matter action $S_m[g]=\int d^4x\,{\cal L}_m[g]$, where ${\cal L}_m[g]$ is the Lagrangian density of matter, is transformed into 

\begin{align} S_m[\hat g]=\int d^4x\,e^{-2\sqrt\frac{2}{2\omega+3}\hat\phi}{\cal L}_m[\hat g],\nonumber\end{align} or, equivalently,

\begin{align} \int d^4x\,{\cal L}_m[g]=\int d^4x\,e^{-2\sqrt\frac{2}{2\omega+3}\hat\phi}{\cal L}_m[\hat g],\label{capoz}\end{align} where matter fields are not explicitly considered. This means that, in the conformal frame, the matter action acquires an anomalous coupling of the matter fields with the scalar $\hat\phi=\int\sqrt{\omega+3/2}\,d\phi/\phi$. 

Another way in which the non-minimal coupling of matter with the scalar field is introduced by the conformal transformation $\hat g_{\mu\nu}=\Omega^2g_{\mu\nu}$, is given by the following equality \cite{sokolowski_prd_1994, kaloper_prd_1998}:

\begin{align} \int d^4x\sqrt{-g}\,L_m(\chi,\der\chi,g_{\mu\nu})=\int d^4x\sqrt{-\hat g}\,\Omega^{-4}L_m(\chi,\hat\der\chi,\Omega^{-2}\hat g_{\mu\nu}),\label{kalo}\end{align} where, as in \eqref{flan-lm-eq} and \eqref{jarv-lm-eq}, the matter fields are not transformed under CTs. From this equality, it follows that not only does the matter Lagrangian acquire a factor $\Omega^{-4}$, but also the matter fields are minimally coupled to the conformal metric $\Omega^{-2}\hat g_{\mu\nu}$ instead of $\hat g_{\mu\nu}$. This has led to the claim that there is only one physical metric: the one to which matter is minimally coupled, that is, the one whose geodesics matter particles follow.

%----------------------------------------

It has been shown that, in contrast to \eqref{flan-lm-eq}, \eqref{jarv-lm-eq}, \eqref{capoz}, and \eqref{kalo}, the matter action \eqref{mac}, is not only conformal invariant $\hat S_m=S_m$, but it is also form-invariant \cite{quiros-2025-a, quiros-2025-b}:

\begin{align} S_m\left[\hat\chi,\hat\der\hat\chi,\hat g_{\mu\nu}\right]=S_m\left[\chi,\der\chi,g_{\mu\nu}\right]\;\Rightarrow\;\int d^4x\sqrt{-\hat g}\,L_m(\hat\chi,\hat\der\hat\chi,\hat g_{\mu\nu})=\int d^4x\sqrt{-g}\,L_m(\chi,\der\chi,g_{\mu\nu}).\label{finv-mac}\end{align} 

This equality entails that if $g_{\mu\nu}$ is the physical metric with respect to the matter fields $\chi$, then $\hat g_{\mu\nu}$ is the physical metric with respect to the matter fields $\hat\chi$. Hence, if the metric $g_{\mu\nu}$ is the physical metric in some frame, its conformal $\hat g_{\mu\nu}=\Omega^2g_{\mu\nu}$ is the physical metric in the conformal frame. This result, obtained based on \eqref{finv-mac}, confronts us with multiple (in principle an infinite number of) physical metrics instead of only one, as it is widely accepted based on \eqref{kalo}. We want to draw attention to the fact that the equations \eqref{flan-lm-eq}, \eqref{jarv-lm-eq}, and \eqref{kalo} are themselves misleading. The fact that the matter fields $\chi$ in these equations are not transformed by conformal transformations means that \eqref{flan-lm-eq}, \eqref{jarv-lm-eq}, and \eqref{kalo} are valid only for null fields. Take, for example, the electromagnetic (EM) radiation field,\footnote{Timelike fields, such as, for example, the fermions, have non-vanishing conformal weight.} where $\chi$ represents the vector potential of the EM field $A_\mu$. It is a well-known fact that null geodesics are not modified by CTs (see Appendix D of \cite{wald-book}). Therefore, if the photons follow the null geodesics of the metric $g_{\mu\nu}$, in the conformal frame, they follow the null geodesics of the conformal metric $\hat g_{\mu\nu}$. This means that, even for radiation, it is \eqref{finv-mac} the correct equality, instead of \eqref{flan-lm-eq}, \eqref{jarv-lm-eq}, or \eqref{kalo}. Therefore, standard matter fields are conformal invariant. To our knowledge, this simple (almost trivial) result has been discussed before only in \cite{quiros-2025-a} in the context of the JFBD parametrization of STG, for the singular value of the BD coupling parameter $\omega=-3/2$.

%-------------------equivalence--------------

%%%%%%%%%%%%%%%%%%%%%%%%%%%%%%%%%%%%%%%%%%%%%%%%%%

\section{Equivalence with Jordan frame Brans-Dicke parametrization: the vacuum case}
\label{sect-equiv-vac}

%%%%%%%%%%%%%%%%%%%%%%%%%%%%%%%%%%%%%%%%%%%%%%%%%%

The gravitational action of a generic STG theory can be written in full resemblance to the JFBD action \cite{brans-dicke-1961, wagoner_1970};

\begin{align} S_\text{jf}=\frac{1}{2}\int d^4x\sqrt{-g}\left[\phi R-\frac{\omega(\phi)}{\phi}(\der\phi)^2-2V(\phi)\right],\label{jfbd-lag}\end{align} where $R$ is the curvature scalar, while the coupling parameter $\omega=\omega(\phi)$ and the self-interaction potential $V=V(\phi),$ are functions of the BD scalar field $\phi$. The BD theory \cite{brans-dicke-1961} is a particular case of \eqref{jfbd-lag} when $\omega(\phi)=\omega_\text{BD},$ is a constant coupling and the self-interaction potential vanishes $V=0$.

Let us demonstrate that the parametrization \eqref{flan-lag}, which is form-invariant under simultaneous transformations \eqref{conf-t} and \eqref{flan-t}, is fully equivalent to the JFBD parametrization \eqref{jfbd-lag}, being form-invariant under the following restricted conformal transformations:\footnote{The transformation of the self-interacting potential in \eqref{gauge-t'}, deserves a clarifying comment. Although from \eqref{gauge-t'} one gets the false impression that the potential can be any function of the BD field $\phi$, as a matter of fact, since $V(\phi)\to\Omega^{-4}V(\phi)$, and $\phi\to\Omega^{-2}\phi$, the only conformal symmetry preserving potential is $V(\phi)=\lambda\,\phi^2$, where $\lambda$ is some dimensionless constant.}

\begin{align} g_{\mu\nu}\rightarrow\hat g_{\mu\nu}=\Omega^2g_{\mu\nu},\;\phi\rightarrow\hat\phi=\Omega^{-2}\phi,\;\chi\rightarrow\hat\chi=\Omega^{w_\chi}\chi,\;\;V\rightarrow\hat V=\Omega^{-4}V,\label{gauge-t'}\end{align} if we add a simultaneous transformation of the coupling parameter $\omega=\omega(\phi)$;

\begin{align} \omega\rightarrow\hat\omega=\frac{\omega+6\frac{\Omega_{,\phi}}{\Omega}\phi\left(1-\frac{\Omega_{,\phi}}{\Omega}\phi\right)}{\left(1-2\frac{\Omega_{,\phi}}{\Omega}\phi\right)^2}.\label{gauge-w-t}\end{align} Note that, if one assigns the specific value $\omega=-3/2$ in \eqref{gauge-w-t}, the numerator of the right-hand side (RHS) can be written as: 

\begin{align} -\frac{3}{2}\left(1-2\frac{\Omega_{,\phi}}{\Omega}\phi\right)^2,\nonumber\end{align} so that $\hat\omega=-3/2$. In the bibliography, this is called the singular value of the coupling parameter, since it is the only value that is not transformed under simultaneous CT \eqref{gauge-t'} and \eqref{gauge-w-t}.

%-------------Demo-------------------------

By redefining a new scalar field and the coupling function in the following way:

\begin{align} \sigma\equiv A(\phi),\;\omega(\sigma)\equiv\frac{A(\phi)B(\phi)}{A'^2(\phi)},\label{w-coup}\end{align} the gravitational piece of the action \eqref{flan-lag} can be written in the following equivalent JFBD form:

\begin{align} S=\frac{1}{2}\int d^4x\sqrt{-g}\left[\sigma R-\frac{\omega}{\sigma}(\der\sigma)^2-2V\right],\label{jfbd-param}\end{align} where $\omega=\omega(\sigma)$ and $V=V(\sigma)$.

Since the scalar field $\phi$ has been removed from the above action, in \eqref{flan-t} we can set $\hat\phi=\phi$, so that $f'=1$. It also follows that

\begin{align} \hat A'=\Omega^{-2}A'\left(1-2\frac{\Omega'}{\Omega}\frac{A}{A'}\right).\label{demo-1}\end{align} Besides, since $\sigma=A(\phi)$, then;

\begin{align} \frac{\Omega'}{A'}=\Omega_{,\sigma}\;\Rightarrow\;\frac{\Omega'}{\Omega}\frac{A}{A'}=\frac{\Omega_{,\sigma}}{\Omega}\,\sigma.\label{demo-2}\end{align} It is not difficult to demonstrate that

\begin{align} \frac{\hat A\hat B}{\hat A'^2}=\frac{\frac{AB}{A'^2}+6\frac{\Omega'}{\Omega}\frac{A}{A'}\left(1-\frac{\Omega'}{\Omega}\frac{A}{A'}\right)}{\left(1-2\frac{\Omega'}{\Omega}\frac{A}{A'}\right)^2},\nonumber\end{align} or, if we take into account \eqref{w-coup}, \eqref{demo-1}, and \eqref{demo-2}, then\footnote{The requirement that the coupling parameter $\omega$ must be transformed simultaneously with CT \eqref{gauge-t'} is expected since, as long as the BD field $\phi$ transforms under CT: $\phi\rightarrow\hat\phi=\Omega^{-2}\phi$, it is clear that any $\phi$-dependent parameter, such as the coupling parameter $\omega=\omega(\phi)$, must also transform. Then, when we consider the restricted CT \eqref{gauge-t'}, we can attribute the fact that \eqref{jfbd-lag} cannot be written in a covariant way with respect to the configuration space, to the unjustified omission of the transformation \eqref{gauge-w-t} of the coupling parameter.}

\begin{align} \hat\omega=\frac{\omega+6\frac{\Omega_{,\sigma}}{\Omega}\sigma\left(1-\frac{\Omega_{,\sigma}}{\Omega}\sigma\right)}{\left(1-2\frac{\Omega_{,\sigma}}{\Omega}\sigma\right)^2}.\nonumber\end{align}

%----------------------------------

Therefore, the transformations \eqref{flan-t} amount to:

\begin{align} \sigma\rightarrow\hat\sigma=\Omega^{-2}\sigma,\;V\rightarrow\hat V=\Omega^{-4}V,\;\omega\rightarrow\hat\omega=\frac{\omega+6\frac{\Omega_{,\sigma}}{\Omega}\,\sigma\left(1-\frac{\Omega_{,\sigma}}{\Omega}\,\sigma\right)}{\left(1-2\frac{\Omega_{,\sigma}}{\Omega}\,\sigma\right)^2}.\label{new-reparam}\end{align} 

In the absence of matter fields, that is, for vacuum, this demonstrates the equivalence of form-invariance of STG theory in the parametrization \eqref{flan-lag} proposed in \cite{flanagan_cqg_2004}, under simultaneous transformations \eqref{conf-t} and \eqref{flan-t}, with form-invariance of STG theory in the JFBD parametrization \eqref{jfbd-lag}, under generalized conformal transformations \eqref{conf-t} and \eqref{new-reparam}. This result was published for the first time in \cite{faraoni_1998} (see also Appendix A of \cite{wands-rev-1999}). 

%----------------------------------

%-------------------equivalence'--------------

%%%%%%%%%%%%%%%%%%%%%%%%%%%%%%%%%%%%%%%%%%%%%%%%%%

\section{Equivalence with Jordan frame Brans-Dicke parametrization: matter fields}
\label{sect-equiv-mat}

%%%%%%%%%%%%%%%%%%%%%%%%%%%%%%%%%%%%%%%%%%%%%%%%%%

The addition of the following matter action:

\begin{align} S_m\left[\chi,\der\chi,g_{\mu\nu}\right]=\int d^4x\sqrt{-g}\,L_m\left(\chi,\der\chi,g_{\mu\nu}\right),\label{mac}\end{align} to the JFBD-type gravitational action \eqref{jfbd-param}, marks the choice of the active approach to the conformal transformations.\footnote{We leave the analysis of the passive approach to CTs for Section \ref{sect-pact}.} This action is form-invariant under CTs \eqref{gauge-t'} plus simultaneous transformation \eqref{mass-t} of the masses (or \eqref{rho-t} in case of perfect fluids with energy density $\rho$) \cite{quiros-2025-a}. The resulting total action,

\begin{align} S_\text{tot}=\frac{1}{2}\int d^4x\sqrt{-g}\left[\sigma R-\frac{\omega}{\sigma}(\der\sigma)^2-2V\right]+\int d^4x\sqrt{-g}\,L_m\left(\chi,\der\chi,g_{\mu\nu}\right),\label{jfbd-tot-action}\end{align} is, consequently, form-invariant under generalized conformal transformations, 

\begin{align} g_{\mu\nu}&\rightarrow\hat g_{\mu\nu}=\Omega^2g_{\mu\nu},\;\sigma\rightarrow\hat\sigma=\Omega^{-2}\sigma,\;\chi\rightarrow\hat\chi=\Omega^{w_\chi}\chi,\;V\rightarrow\hat V=\Omega^{-4}V,\nonumber\\
\omega&\rightarrow\hat\omega=\frac{\omega+6\frac{\Omega_{,\sigma}}{\Omega}\,\sigma\left(1-\frac{\Omega_{,\sigma}}{\Omega}\,\sigma\right)}{\left(1-2\frac{\Omega_{,\sigma}}{\Omega}\,\sigma\right)^2},\;m\to\hat m=\Omega^{-1}m\;\;(\rho\to\hat\rho=\Omega^{-4}\rho).\label{gct}\end{align} This generalizes the result of \cite{faraoni_1998} for vacuum, by including the coupling of arbitrary matter fields to gravity. 

In view of the equivalence demonstrated above, we can say that STG theory in the parametrization \eqref{flan-lag};

\begin{align} S_\text{tot}=\frac{1}{2}\int d^4x\sqrt{-g}\left[AR-B(\der\phi)^2-2V\right]+\int d^4x\sqrt{-g}\,L_m\left(\chi,\der\chi,g_{\mu\nu}\right),\label{flan-lag'}\end{align} which is equivalent to \eqref{jfbd-tot-action}, is form-invariant under \eqref{flan-t} plus simultaneous conformal transformations,

\begin{align} \hat g_{\mu\nu}=\Omega^2g_{\mu\nu},\;\hat\chi=\Omega^{\omega_\chi}\chi,\;\hat m=\Omega^{-1}m\;\;(\hat\rho=\Omega^{-4}\rho).\label{gen-flan-t}\end{align} 

Note that in \eqref{flan-lag'}, we have replaced the matter action \eqref{flan-mat-action} proposed in \cite{flanagan_cqg_2004}, where the matter fields are minimally coupled to the metric $e^{2\alpha}g_{\mu\nu}$, by the conformal form-invariant matter action \eqref{mac}, where the matter fields (including timelike fields) are minimally coupled to the metric $g_{\mu\nu}$.

%-------------EOM--------------

Conformal form-invariance of the matter action in \eqref{flan-lag'} leads to the following Ward identity \cite{quiros-2025-b} (see Appendix \ref{app-a}):

\begin{align} \frac{\delta{\cal L}_m}{\delta\phi}=\frac{A'}{2A}\,\sqrt{-g}\,T^{(m)}.\label{ward-id}\end{align} This equation plays an important role in the derivation of the correct equations of motion when conformal invariance is a symmetry of the matter Lagrangian density. 

The correct (conformal form-invariant) EOMs, which are obtained through the variational procedure from \eqref{flan-lag'}, taking into account \eqref{ward-id}, are the following:

\begin{align} A{\cal E}_{\mu\nu}&=T^{(m)}_{\mu\nu},\label{einst-eom}\\
2AB\nabla^2\phi+AB'(\der\phi)^2+A'\left(AR+T^{(m)}\right)&=2AV',\label{jfkg-eom}\end{align} where

\begin{align} {\cal E}_{\mu\nu}:=&G_{\mu\nu}-\left(\frac{A''+B}{A}\right)\der_\mu\phi\der_\nu\phi+\left(\frac{2A''+B}{2A}\right)(\der\phi)^2g_{\mu\nu}-\frac{A'}{A}\left(\nabla_\mu\nabla_\nu-g_{\mu\nu}\nabla^2\right)\phi+\frac{V}{A}\,g_{\mu\nu},\label{E-def}\\
{\cal E}\equiv&g^{\mu\nu}{\cal E}_{\mu\nu}=-R+\left(\frac{3A''+B}{A}\right)(\der\phi)^2+3\frac{A'}{A}\nabla^2\phi+4\frac{V}{A}.\label{E-trace}\end{align} The trace of the Einstein-type equation \eqref{einst-eom} can be written as;

\begin{align} 3A'\nabla^2\phi+(3A''+B)(\der\phi)^2+4V=AR+T^{(m)}.\label{trace-einst-eom}\end{align} With the help of this equation, we can eliminate the combination $AR+T^{(m)}$ from the JFKG-EOM \eqref{jfkg-eom}. We get that,

\begin{align} \left(2AB+3A'^2\right)\nabla^2\phi+\frac{1}{2}\left(2AB+3A'^2\right)'(\der\phi)^2=2\left(AV'-2A'V\right),\nonumber\end{align} or, if we choose the only self-interaction potential that preserves conformal symmetry: $V=\kappa\,A^2$, then $AV'=2A'V$. That is, the KG-EOM reads as,

\begin{align} \left(2AB+3A'^2\right)\nabla^2\phi+\frac{1}{2}\left(2AB+3A'^2\right)'(\der\phi)^2=0\;\Rightarrow\;\nabla^2\phi+\beta(\phi)(\der\phi)^2=0,\label{phi-eom}\end{align} where the parameter $\beta=\beta(\phi)\equiv d\ln\sqrt{2AB+3A'^2}/d\phi$. This KG-type EOM is independent of the matter content and of the curvature of spacetime. That is, it does not affect the gravitational dynamics of matter. Only the choice of the parameters $A=A(\phi)$ and $B=B(\phi)$ has consequences for the solution $\phi=\phi(x)$.

This result has no parallel in the bibliography on STG theory in the JFBD parametrization. In the standard bibliography, the conformal form-invariance of the matter action is missed out, so the Ward identity \eqref{ward-id} is omitted. The resultant KG-type EOM, which is derived by varying the action \eqref{flan-lag'}, reads:

\begin{align} 2B\nabla^2\phi+B'(\der\phi)^2+A'R-2V'=0.\label{wrong-jfkg-eom}\end{align} In this case, the trace equation \eqref{trace-einst-eom} is used to eliminate $R$ from the KG-EOM. Actually, if we substitute $R$ from \eqref{trace-einst-eom} into \eqref{wrong-jfkg-eom}, we get,

\begin{align} \nabla^2\phi+\beta(\phi)(\der\phi)^2=\frac{2(AV'-2A'V)}{2AB+3A'^2}+\frac{A'\,T^{(m)}}{2AB+3A'^2}.\label{wrong-jfkg-eom'}\end{align} This is indeed a dynamical EOM. When $A=\phi$ and $B=\omega(\phi)/\phi$, this EOM reduces to the well-known JFKG-EOM;

\begin{align} \nabla^2\phi+\frac{\omega'}{2\omega+3}(\der\phi)^2=\frac{2(\phi V'-2V)}{2\omega+3}+\frac{T^{(m)}}{2\omega+3},\nonumber\end{align} which is incorrect unless $T^{(m)}=0$, or if one can just omit at will a symmetry of the action and of the derived EOM (in the present case conformal symmetry).

%==========================================

\subsection{Conservation of stress-energy}

%==========================================

Let us take the divergence of the Einstein-type EOM \eqref{einst-eom}: $\nabla^\lambda(A{\cal E}_{\lambda\mu})=\nabla^\lambda T^{(m)}_{\lambda\mu}$, and use the second Bianchi identity $\nabla^\lambda G_{\lambda\mu}=0$, and the identity $(\nabla_\lambda\nabla_\mu-\nabla_\mu\nabla_\lambda)\nabla^\lambda\phi=R_{\mu\lambda}\nabla^\lambda\phi$. We get

\begin{align} \nabla^\lambda T^{(m)}_{\lambda\mu}=-\frac{1}{2}\nabla_\mu\phi\left[2B\nabla^2\phi+B'(\der\phi)^2+A'R-2V'\right].\label{set-eom}\end{align} If in the right-hand side (RHS) of \eqref{set-eom} we substitute the correct JFKG-EOM if the Lagrangian density of matter is conformal form-invariant, i.e., equation \eqref{jfkg-eom}, that can be written in the following way:

\begin{align} 2B\nabla^2\phi+B'(\der\phi)^2+A'R-2V'=-\frac{A'}{A}\,T^{(m)},\nonumber\end{align} we obtain the following nonhomogeneous continuity equation \cite{quiros-2025-b}:

\begin{align} \nabla^\lambda T^{(m)}_{\lambda\mu}=\frac{A'}{2A}\nabla_\mu\phi\,T^{(m)}.\label{set-eom'}\end{align} 

Note that, if we just forget about conformal form-invariance of the Lagrangian density of matter and we omit the Ward identity \eqref{ward-id}; that is, if we substitute \eqref{wrong-jfkg-eom} in \eqref{set-eom}, we obtain the standard conservation equation $\nabla^\lambda T^{(m)}_{\lambda\mu}=0$.

The nonvanishing RHS of \eqref{set-eom'} amounts to a fifth-force. This is required for conformal form-invariance of \eqref{set-eom'} to hold. Recall that the standard conservation equation $\nabla^\lambda T^{(m)}_{\lambda\mu}=0$, is conformal form-invariant only for radiation, but not in general. The arising fifth-force acts only on timelike fields. It does not act on radiation and null fields in general, since $T^{(m)}=0$. This is an encouraging result for conformal symmetry, since the fifth-force is a kind of ``dark force'', which can be investigated as an alternative to explain, in principle, the dark cosmological sector.

%--------------------------------------------

%%%%%%%%%%%%%%%%%%%%%%%%%%%%%%%%%%%%%%%%%%%%%%%%%%%%%%%%%%

\section{Mathematical basis of passive and active conformal transformations}
\label{sect-math-base}

%%%%%%%%%%%%%%%%%%%%%%%%%%%%%%%%%%%%%%%%%%%%%%%%%%%%%%%%%%

If we adopt the passive approach to CTs, the inclusion of any matter field in the formalism proposed in \cite{flanagan_cqg_2004} requires replacement of the matter action \eqref{flan-mat-action} by (see equations \eqref{cinv-dirac-action} and \eqref{cinv-proca-action} in Appendix \ref{app-c}),

\begin{align} S_m\left[\Psi,\mathfrak{D}\Psi,\mathfrak{g}_{\mu\nu}\right]=\int d^4x\sqrt{-\mathfrak{g}}\,L_m\left(\Psi,\mathfrak{D}\Psi,\mathfrak{g}_{\mu\nu}\right),\label{mat-action}\end{align} where, under the conformal transformations, $\Psi\rightarrow\hat\Psi=\Psi$ and $\mathfrak{g}_{\mu\nu}\rightarrow\hat{\mathfrak{g}}_{\mu\nu}=\mathfrak{g}_{\mu\nu}$. This means that CTs \eqref{gauge-t} acting on $g_{\mu\nu}$, $\phi$ and $\chi$ become the identity transformation of the physical fields $\mathfrak{g}_{\mu\nu}$ and $\Psi$. Therefore, conformal form-invariance is not an actual symmetry of the matter action \eqref{mat-action}. This can be seen more easily if we notice that the auxiliary fields $\phi$, $\chi$, and $g_{\mu\nu}$, which suffer the conformal transformations \eqref{gauge-t}, do not explicitly appear in \eqref{mat-action}, that is, the latter is written in terms of the conformal invariant fields $\Psi$ and $\mathfrak{g}_{\mu\nu}$ alone. We can say that the matter action \eqref{mat-action} is written in a covariant way in the configuration space and that the conformal symmetry is a fictitious symmetry.

Alternatively, one may write the conformal form-invariant matter action in terms of the matter fields $\chi$ which are minimally coupled to the metric $g_{\mu\nu}$, as defined in \eqref{mac}:

\begin{align} S_m\left[\chi,\der\chi,g_{\mu\nu}\right]=\int d^4x\sqrt{-g}\,L_m\left(\chi,\der\chi,g_{\mu\nu}\right).\nonumber\end{align} 

%----------------------------------------------

A straightforward demonstration that \eqref{mac} and \eqref{mat-action} are equivalent, i. e., 

\begin{align} S_m\left[\chi,\der\chi,g_{\mu\nu}\right]\equiv S_m\left[\Psi,\mathfrak{D}\Psi,\mathfrak{g}_{\mu\nu}\right],\label{equiv-mat-action}\end{align} is given in \cite{quiros-2025-a, quiros-2025-b}. A specific illustration is given by the following equality (see Appendix \ref{app-c}): 

\begin{align} \int d^4x\sqrt{-g}\,\bar\psi\left(i\cancel{\cal D}+m\right)\psi\equiv\int d^4x\sqrt{-\mathfrak{g}}\,\bar\Psi\left(i\cancel{\mathfrak{D}}+\mathfrak{m}\right)\Psi.\label{equiv-demo}\end{align} For the matter action of a perfect fluid the equivalence \eqref{equiv-mat-action} can be written in the following way: 

\begin{align} \int d^4x\sqrt{-g}\,\rho\equiv\int d^4x\sqrt{-\mathfrak{g}}\,\mathfrak{r},\label{pfluid-equiv}\end{align} where we choose the Lagrangian of the perfect fluid $L_\text{fluid}=-\rho$ \cite{hawking-book, faraoni_2009, berto-prd-2008}, with pressure $p$ and energy density $\rho$ related by the equation of state (EOS) $p=w\rho$ (the number $w$ is the EOS parameter of the fluid). In the RHS of \eqref{pfluid-equiv} $\mathfrak{r}:=e^{-4\alpha}\rho$ is the conformal invariant energy density of the perfect fluid, which is related to the conformal invariant pressure of the fluid by $\mathfrak{p}=w\,\mathfrak{r}$. A similar equivalence for perfect fluids, but with different conformal invariant (physical) metric and energy density, has been demonstrated in \cite{quiros-2025-b}.

%---------------------------------

Conformal invariance is a manifest symmetry of the LHS of equivalence \eqref{equiv-mat-action} because, under CT: $\hat g_{\mu\nu}=\Omega^2g_{\mu\nu}$, $\hat\chi=\Omega^{w_\chi}\chi$, the matter action \eqref{mac} is not only conformal invariant, but also form-invariant;

\begin{align} \int d^4x\sqrt{-g}\,L_m\left(\chi,\der\chi,g_{\mu\nu}\right)=\int d^4x\sqrt{-\hat g}\,L_m\left(\hat\chi,\hat\der\hat\chi,\hat g_{\mu\nu}\right)\;\Rightarrow\;S_m[\chi,\der\chi,g_{\mu\nu}]=S_m[\hat\chi,\hat\der\hat\chi,\hat g_{\mu\nu}].\nonumber\end{align} Meanwhile, the RHS of \eqref{equiv-mat-action}, is written in terms of conformal invariant fields $\mathfrak{g}_{\mu\nu}$ and $\Psi$, which are not transformed by CTs. Hence, the conformal form-invariance is not an actual symmetry of the matter action \eqref{mat-action}

%---------------------------------

The passive approach to CT must be based on conformal invariant fields and operators. The matter action defines the conformal invariant metric to which the conformal invariant matter fields are minimally coupled; that is, it defines the physical metric and physical matter fields. Due to the equivalence \eqref{equiv-mat-action}, the total action \eqref{flan-lag'} can be written in the following way: 

\begin{align} S_\text{tot}=\frac{1}{2}\int d^4x\sqrt{-g}\left[AR-B(\der\phi)^2-2V\right]+\int d^4x\sqrt{-\mathfrak{g}}\,L_m\left(\Psi,\mathfrak{D}\Psi,\mathfrak{g}_{\mu\nu}\right),\label{arb}\end{align} where the gravitational action must be written in terms of the physical metric \eqref{phys-g}: $\mathfrak{g}_{\mu\nu}=e^{2\alpha}g_{\mu\nu}$ and, perhaps, other conformal invariant fields (see below). We want to underline that there are many ways in which the conformal invariant metric can be defined, for example, in \cite{jarv_2015} the conformal invariant metric $g^{(A)}_{\mu\nu}\equiv A(\phi)g_{\mu\nu}$ was chosen (see Section \ref{sect-inv-q}). As long as the matter action is given by \eqref{mat-action} with $\mathfrak{g}_{\mu\nu}=e^{2\alpha}g_{\mu\nu}$, the conformal invariant metric $g^{(A)}_{\mu\nu}$ is not the physical metric, since it is not the metric to which the matter fields are minimally coupled. For a very interesting and enlightening discussion of this subject, we submit the reader to \cite{brans-1988}. Once the total action \eqref{arb} is written in terms of only conformal invariant fields and operators, that is, once it is written in covariant form in the configuration space, the fields $g_{\mu\nu}$, $\phi$, and $\chi$, which suffer the conformal transformations \eqref{gauge-t}, become auxiliary fields that are removed from the covariant description. Different sets of fields $\{g_{\mu\nu},\phi,\chi\}$ and $\{\hat g_{\mu\nu},\hat\phi,\hat\chi\}$ provide different representations of the same gravitational state ${\cal S}_\mathfrak{g}:(\Psi,\mathfrak{g}_{\mu\nu})$ (see the figure on the left in FIG. \ref{fig1}).

%-----------------------------------------

In the framework of STG theory in JFBD parametrization, in the configuration space ${\cal M}_\text{fields}$, active CT should be based on the following total action:

\begin{align} S_\text{tot}=\frac{1}{2}\int d^4x\sqrt{-g}\left[\phi R-\frac{\omega}{\phi}(\der\phi)^2-2V\right]+\int d^4x\sqrt{-g}\,L_m(\chi,\der\chi,g_{\mu\nu}),\label{tot-action}\end{align} where $\omega=\omega(\phi)$, $V=V(\phi)$. Here, the metric $g_{\mu\nu}$ and the matter fields $\chi$, are physical fields, while the BD scalar $\phi$ is a gauge field. The total action \eqref{tot-action} and the derived EOM are form-invariant under simultaneous CT \eqref{gauge-t'} and \eqref{gauge-w-t}, being a generalization of the conformal form-invariant vacuum JFBD theory \eqref{jfbd-lag} studied in \cite{faraoni_1998}, to the presence of any matter fields. In the present case, the conformal transformations relate different gravitational states ${\cal S}_g:\left(\chi,\phi,g_{\mu\nu}\right)$ and $\hat{\cal S}_g:\left(\hat\chi,\hat\phi,\hat g_{\mu\nu}\right)$ (see the figure on the right in Fig. \ref{fig1}).

%----------------------------------

%-----------------------------------

%%%%%%%%%%%%%%%%%%%%%%%%%%%%%%%%%%%%%%%%%%%%%%%%%%%%%%%%

\section{Passive conformal transformations and the form-invariant parametrization}
\label{sect-pact}

%%%%%%%%%%%%%%%%%%%%%%%%%%%%%%%%%%%%%%%%%%%%%%%%%%%%%%%%

To get a different perspective on the form-invariant parametrization proposed in \cite{flanagan_cqg_2004}, we shall write the gravitational action \eqref{flan-lag} in terms of the explicitly declared physical metric $\mathfrak{g}_{\mu\nu}\equiv e^{2\alpha(\phi)}g_{\mu\nu}$, which is the metric that defines the proper time measured by atomic clocks and to which the matter fields are minimally coupled. Our goal is to write the action \eqref{flan-lag} in a covariant way in the configuration space manifold. For this purpose, let us consider the following conformal invariant (physical) quantities:

\begin{align} \mathfrak{g}_{\mu\nu}=e^{2\alpha}g_{\mu\nu},\;\vphi=e^{-2\alpha}A,\;\Psi=e^{w_\chi\alpha}\chi.\label{phys-g-vphi}\end{align} In terms of the above fields, the gravitational action in \eqref{flan-lag} (same as in \eqref{arb}), transforms into,

\begin{align} S_\text{grav}=\frac{1}{2}\int d^4x\sqrt{-\mathfrak{g}}\left[\vphi\mathfrak{R}-{\cal W}\frac{(\mathfrak{V}\vphi)^2}{\vphi}-2{\cal V}\right],\label{g-phys-jarv-action}\end{align} where the physical curvature scalar $\mathfrak{R}$ and the covariant derivative $\mathfrak{V}_\mu$ are defined with respect to the physical metric $\mathfrak{g}_{\mu\nu}$: $(\mathfrak{V}\vphi)^2\equiv\mathfrak{g}^{\mu\nu}\mathfrak{V}_\mu\vphi\mathfrak{V}_\nu\vphi$, while ${\cal V}=e^{-4\alpha}V$ is the form-invariant self-interaction potential and ${\cal W}={\cal W}(\vphi)$ (see Appendix \ref{app-b} for details and notation);

\begin{align} {\cal W}=\frac{B/A-6\alpha'(\alpha'-A'/A)}{\left(2\alpha'-A'/A\right)^2},\label{phys-w}\end{align} is the form-invariant coupling function. To obtain \eqref{g-phys-jarv-action}, we took into account that up to a total divergence:

\begin{align} \sqrt{-g}\,A(\phi)R=\sqrt{-\mathfrak{g}}\,Ae^{-2\alpha}\left[\mathfrak{R}+6\alpha'\left(\alpha'-\frac{A'}{A}\right)(\mathfrak{D}\phi)^2\right],\nonumber\end{align} and that

\begin{align} Ae^{-2\alpha}(\mathfrak{D}\phi)^2=\frac{1}{\left(2\alpha'-\frac{A'}{A}\right)^2}\frac{(\mathfrak{D}\vphi)^2}{\vphi}.\nonumber\end{align} 

The next step is to add the matter action \eqref{mat-action}:

\begin{align} S_m=\int d^4x\sqrt{-\mathfrak{g}}\,L_m\left(\Psi,\mathfrak{D}\Psi,\mathfrak{g}_{\mu\nu}\right),\nonumber\end{align} where $L_m\left(\Psi,\mathfrak{D}\Psi,\mathfrak{g}_{\mu\nu}\right)$ is the Lagrangian of the conformal invariant matter fields $\Psi$, which are coupled to the conformal invariant metric $\mathfrak{g}_{\mu\nu}$, to the gravitational action \eqref{g-phys-jarv-action}. The total action of the resulting theory reads:

\begin{align} S_\text{tot}=\frac{1}{2}\int d^4x\sqrt{-\mathfrak{g}}\left[\vphi\mathfrak{R}-{\cal W}\frac{(\mathfrak{V}\vphi)^2}{\vphi}-2{\cal V}\right]+\int d^4x\sqrt{-\mathfrak{g}}\,L_m\left(\Psi,\mathfrak{D}\Psi,\mathfrak{g}_{\mu\nu}\right).\label{phys-flan-lag}\end{align} This action is the same as in \eqref{arb}, but with the gravitational piece written in terms of physical (conformal invariant) quantities. It is also equivalent to the JFBD-type action \eqref{tot-action} for a generic STG theory, with the difference that the application of a conformal transformation to \eqref{tot-action} means the identity transformation in \eqref{phys-flan-lag}. 

%-----------------------------------

It is evident that neither the conformal invariant scalar field $\vphi$, the physical metric $\mathfrak{g}_{\mu\nu}$ and the derived quantities such as $\mathfrak{R}$ and $(\mathfrak{D}\vphi)^2/\vphi$, nor the conformal invariant self-interaction potential ${\cal V}$, are transformed by generalized CT \eqref{conf-t} and \eqref{flan-t}. The same is true for the Lagrangian density of matter ${\cal L}_m=\sqrt{-\mathfrak{g}}\,L_m\left(\Psi,\mathfrak{D}\Psi,\mathfrak{g}_{\mu\nu}\right)$. Hence, it remains to show that the coupling function ${\cal W}$ is form-invariant under these transformations. This is easily shown if we realize that under \eqref{flan-t}:

\begin{align} \hat\alpha'=(f')^{-1}\left(\alpha'-\frac{\Omega'}{\Omega}\right),\;\frac{\hat A'}{\hat A}=(f')^{-1}\left(\frac{A'}{A}-2\frac{\Omega'}{\Omega}\right),\;\frac{\hat B}{\hat A}=(f')^{-2}\left[\frac{B}{A}+6\frac{\Omega'}{\Omega}\frac{A'}{A}-6\left(\frac{\Omega'}{\Omega}\right)^2\right].\label{usef-t}\end{align} From the above equations, it follows that

\begin{align} \left(2\hat\alpha'-\frac{\hat A'}{\hat A}\right)^2=\left(f'\right)^{-2}\left(2\alpha'-\frac{A'}{A}\right)^2,\;\frac{\hat B}{\hat A}-6\hat\alpha'\left(\hat\alpha'-\frac{\hat A'}{\hat A}\right)=\left(f'\right)^{-2}\left[\frac{B}{A}-6\alpha'\left(\alpha'-\frac{A'}{A}\right)\right].\nonumber\end{align} Hence, it is verified that $\hat{\cal W}={\cal W}$. That is, the coupling function \eqref{phys-w} is not transformed. This completes the demonstration that the formalism proposed in \cite{flanagan_cqg_2004} can be written in a form-invariant way \eqref{phys-flan-lag} in the configuration space and that the CTs amount to the identity transformation of \eqref{phys-flan-lag}. 

%---------------------------------------------

Notice that the auxiliary fields $g_{\mu\nu}$ and $\phi$, as well as the field-dependent parameters $\alpha=\alpha(\phi)$ and $A=A(\phi)$, have been completely removed from the covariant description \eqref{phys-flan-lag} under PACT. This means that the generalized CTs \eqref{conf-t} and \eqref{flan-t} have been replaced by the identity transformation.\footnote{Under the conformal transformation $g_{\mu\nu}\rightarrow\Omega^2g_{\mu\nu}$, the physically meaningful quantities in \eqref{phys-flan-lag} are not transformed: $\mathfrak{g}_{\mu\nu}\rightarrow\mathfrak{g}_{\mu\nu}$, $\mathfrak{R}\rightarrow\mathfrak{R}$, $\vphi\rightarrow\vphi$, $\Psi\rightarrow\Psi,$ $\mathfrak{V}_\mu\vphi\rightarrow\mathfrak{V}_\mu\vphi$, $\mathfrak{V}_\mu\Psi\rightarrow\mathfrak{V}_\mu\Psi$, ${\cal W}\rightarrow{\cal W}$, ${\cal V}\rightarrow{\cal V}.$} In other words: conformal symmetry is hidden behind the identity transformations in the covariant description. Hence, conformal form-invariance is not a Noether symmetry of the theory \eqref{phys-flan-lag}, which is fully equivalent to \eqref{flan-lag}. That is, if we adopt PACT, conformal symmetry can only be a spurious or fictitious symmetry of STG theory in the parametrization proposed in \eqref{phys-flan-lag}. We find that the analysis of conformal symmetry within the passive approach to CTs is not a well-posed problem.

%--------------------------------------

The present analysis has limitations. It is known that in the low-energy, small-curvature limit of string theory, different matter fields minimally couple to different conformal metrics \cite{wands-rev-1999, khoury-prl-2004, khoury-prd-2004}. In this case, the problem is which of the several conformal metrics that are minimally coupled to different matter fields is the one to identify as ``the physical metric.'' Although this issue cannot have a simple solution, in this case we are not sure that conformal symmetry survives at the large energies implied by the string effective theory.

%--------------------------------------

%%%%%%%%%%%%%%%%%%%%%%%%%%%%%%%%%%%%%%%%%%%%

\section{Invariant quantities in STG}
\label{sect-inv-q}

%%%%%%%%%%%%%%%%%%%%%%%%%%%%%%%%%%%%%%%%%%%%

In \cite{jarv_2015} a scheme was proposed to construct conformal invariant quantities starting from the parametrization proposed in \cite{flanagan_cqg_2004}. Among these, we have the following:

\begin{align} &I_1:=\frac{e^{2\alpha}}{A},\;I_2:=\frac{V}{A^2},\;dI_3:=\pm\sqrt{F}d\phi,\label{jarv-inv}\end{align} where

\begin{align} F=\frac{2AB+3A'^2}{4A^2}.\label{jarv-f}\end{align} In that bibliographic reference, it was discussed how to formulate the theory \eqref{flan-lag} in terms of these invariants. The resulting action is

\begin{align} S=\frac{1}{2}\int d^4x\sqrt{-g_{(A)}}\left[R_{(A)}-2(\der_{(A)}I_3)^2-2I_2\right]+S_m\left[\chi,I_1g^{(A)}_{\mu\nu}\right],\label{jarv-action}\end{align} where the quantities and operators with the label ``$(A)$'' are defined with respect to the conformal invariant metric $g^{(A)}_{\mu\nu}\equiv A\,g_{\mu\nu}$, and we have used the following notation: $(\der_{(A)}I_3)^2\equiv g^{\mu\nu}_{(A)}\der_\mu I_3\der_\nu I_3$. Notice that in \eqref{jarv-action}:

\begin{align} I_1\,g^{(A)}_{\mu\nu}=I_1\,A\,g_{\mu\nu}=e^{2\alpha}\,g_{\mu\nu}=\mathfrak{g}_{\mu\nu},\label{phys-met}\end{align} coincides with the physical (conformal form-invariant) metric in \eqref{phys-g-vphi}, so that 

\begin{align} S_m\left[\chi,I_1g^{(A)}_{\mu\nu}\right]=S_m\left[\chi,e^{2\alpha}\,g_{\mu\nu}\right]=S_m\left[\chi,\mathfrak{g}_{\mu\nu}\right],\label{sm-flan}\end{align} which is the matter action proposed in \cite{flanagan_cqg_2004}. As we have shown in Section \ref{sect-math-base}, in the general case, that is, for matter fields different from radiation, this matter action is not conformal form-invariant.

 A problem with theories of the kind \eqref{flan-lag} and its equivalent \eqref{jarv-action} is that there are several different conformal invariant metrics: the gravitational metric $g^{(A)}_{\mu\nu}$ and the ``physical metric'' $\mathfrak{g}_{\mu\nu}$, among other possibilities. As stated in \cite{flanagan_cqg_2004}, the proper time associated with $\mathfrak{g}_{\mu\nu}$ is the time measured by the clocks. In this regard, it is useful to bring to our attention the discussion on this topic in \cite{brans-1988}. It is stated in the latter bibliographic reference that the metric tensor has macroscopic measurable significance, which is determined by its interaction with matter: the form of the matter Lagrangian determines what the interaction between the metric and matter fields really is. This is the only way in which classical fields can have operational meaning. In addition to this, as also stated in \cite{brans-1988}, the operational significance of the metric is found in the spacetime geometry, which is revealed by classical particle paths. These are required to be geodesic paths according to the equivalence principle. Given that the point particles and fields in \eqref{flan-lag} or \eqref{jarv-action} follow geodesics of the physical metric $\mathfrak{g}_{\mu\nu}$, it makes sense to renounce to an additional ``gravitational metric'' and write the overall action \eqref{jarv-action} in terms of the physical metric alone. 

In order to write the action \eqref{jarv-action} in terms of the physical metric \eqref{phys-met}, one has to make the following substitution everywhere:

\begin{align} g^{(A)}_{\mu\nu}&\rightarrow\mathfrak{g}_{\mu\nu}=I_1\,g^{(A)}_{\mu\nu}\Rightarrow\,g^{(A)}_{\mu\nu}=\vphi\,\mathfrak{g}_{\mu\nu},\label{replac}\end{align} where we have defined the conformal invariant scalar $\vphi:=I_1^{-1}$. In addition, we have to take into account the following definitions and equations:

\begin{itemize}

    \item The LC connection of the metric $g^{(A)}_{\mu\nu}$ and its relationship with the LC connection of the physical metric $\mathfrak{g}_{\mu\nu}$:

    \begin{align} ^{(A)}\Gamma^\alpha_{\mu\nu}=\frac{1}{2}g^{\alpha\lambda}_{(A)}\left[\der_\nu g^{(A)}_{\mu\lambda}+\der_\mu g^{(A)}_{\nu\lambda}-\der_\lambda g^{(A)}_{\mu\nu}\right]=\mathfrak{G}^\alpha_{\mu\nu}+\mathfrak{L}^\alpha_{\mu\nu},\label{lc-c-rel}\end{align} where the LC affine connection of the physical metric is defined in \eqref{phys-lc-c} and

    \begin{align} \mathfrak{L}^\alpha_{\mu\nu}\equiv\frac{1}{2\vphi}\left[\delta^\alpha_\mu\der_\nu\vphi+\delta^\alpha_\nu\der_\mu\vphi-\mathfrak{g}_{\mu\nu}\mathfrak{g}^{\alpha\lambda}\der_\lambda\vphi\right].\nonumber\end{align}
    
    \item Relationship between the scalar densities:

    \begin{align} \sqrt{-g_{(A)}}R_{(A)}=\sqrt{-\mathfrak{g}}\left[\vphi\,\mathfrak{R}+\frac{3}{2\vphi}(\mathfrak{V}\vphi)^2\right],\label{curv-rel}\end{align} where quantities and operators with the label ``$(A)$'' are defined with respect to the conformal invariant metric $g^{(A)}_{\mu\nu}$ while the quantities and operators in the RHS of the equation are defined with respect to the physical metric. In addition, we have omitted a covariant divergence $-\sqrt{-\mathfrak{g}}\,3\mathfrak{V}^2\vphi$, which amounts to a boundary term in the action. Here we have used the following notation: $(\mathfrak{V}\vphi)^2\equiv\mathfrak{g}^{\mu\nu}\mathfrak{V}_\mu\vphi\mathfrak{V}_\nu\vphi$ and $\mathfrak{V}^2\equiv\mathfrak{g}^{\mu\nu}\mathfrak{V}_\mu\mathfrak{V}_\nu$, where $\mathfrak{V}_\mu$ is the covariant derivative operator which is defined with respect to the physical LC affine connection $\mathfrak{G}^\alpha_{\mu\nu}$.
    
\end{itemize} After the above specifications, it is not difficult to demonstrate that

\begin{align} \sqrt{-g_{(A)}}\left[R_{(A)}-2(\der_{(A)}I_3)^2\right]=\sqrt{-\mathfrak{g}}\left[\vphi\mathfrak{R}-{\cal W}\frac{(\mathfrak{V}\vphi)^2}{\vphi}\right],\nonumber\end{align} where the coupling function ${\cal W}$ is defined in \eqref{phys-w};

\begin{align} {\cal W}=\frac{B/A-6\alpha'(\alpha'-A'/A)}{\left(2\alpha'-A'/A\right)^2}.\nonumber\end{align} Finally, if we take into account that

\begin{align} \sqrt{-g_{(A)}}\,I_2=\sqrt{-\mathfrak{g}}\,\vphi^2\frac{V}{A^2}=\sqrt{-\mathfrak{g}}\,e^{-4\alpha}V=\sqrt{-\mathfrak{g}}\,{\cal V},\nonumber\end{align} it is shown that, written in terms of the physical metric, the gravitational piece of action \eqref{jarv-action} transforms into:

\begin{align} S=\frac{1}{2}\int d^4x\sqrt{-\mathfrak{g}}\left[\vphi\mathfrak{R}-{\cal W}\frac{(\mathfrak{V}\vphi)^2}{\vphi}-2{\cal V}\right],\nonumber\end{align} which coincides with \eqref{g-phys-jarv-action}, as expected, since \eqref{flan-lag} and \eqref{jarv-action} are equivalent parametrizations. Therefore, no matter what parametrization \eqref{flan-lag} or \eqref{jarv-action} we start with, once we identify a single physical metric to account for the gravitational interactions of matter, the resulting STG theory is given by the JFBD parametrization \eqref{phys-flan-lag}, which represents the covariant formulation of STG theory in the configuration space according to the passive approach to CTs. It is not suitable for the investigation of conformal symmetry, since conformal transformations of the auxiliary fields $g_{\mu\nu}$, $\phi$, and $\chi$, amount to the identity transformation of the physical fields $\mathfrak{g}_{\mu\nu}$ and $\Psi$.

%-------------------------------------------

%======================================================

\subsection{A different choice of the physical metric}
\label{sect-alt-phmet}

%======================================================

In the previous analysis, we have kept an element of the proposal in \cite{flanagan_cqg_2004} for the matter action: the metric that is measured by atomic clocks is the conformal invariant metric $\mathfrak{g}_{\mu\nu}=e^{2\alpha(\phi)}g_{\mu\nu}$. However, there is no need to introduce an additional parameter $\alpha=\alpha(\phi)$, to have a well-defined physical metric. In fact, we can safely remove this parameter, and yet we can define a conformal invariant metric that is measured by the clocks. In \cite{quiros-2025-b}, for example, the following definition of the conformal invariant metric and matter fields was proposed:

\begin{align} \hat{\mathfrak{g}}_{\mu\nu}=\frac{\phi}{M^2_\text{pl}}\,g_{\mu\nu},\;\hat\Psi=\left(\frac{\phi}{M^2_\text{pl}}\right)^\frac{w_\chi}{2}\chi,\label{phys-met-mat}\end{align} where $\phi$ is the BD scalar field in the gravitational action in \eqref{tot-action}. Defining the Christoffel symbols in terms of the conformal invariant metric $\hat{\mathfrak{g}}_{\mu\nu}$ in \eqref{phys-met-mat};

\begin{align} \hat{\mathfrak{C}}^\alpha_{\;\;\mu\nu}:=\frac{1}{2}\hat{\mathfrak{g}}^{\alpha\lambda}\left(\der_\nu\hat{\mathfrak{g}}_{\mu\lambda}+\der_\mu\hat{\mathfrak{g}}_{\nu\lambda}-\der_\lambda\hat{\mathfrak{g}}_{\mu\nu}\right),\nonumber\end{align} and the related Riemann-Christoffel curvature tensor $\hat{\mathfrak{R}}^\alpha_{\;\;\mu\beta\nu}$ and its contractions $\hat{\mathfrak{R}}_{\mu\nu}=\hat{\mathfrak{g}}^{\lambda\sigma}\hat{\mathfrak{R}}_{\lambda\mu\sigma\nu}$ and $\hat{\mathfrak{R}}=\hat{\mathfrak{g}}^{\mu\nu}\hat{\mathfrak{R}}_{\mu\nu}$, we can obtain the relationships between these quantities and those defined with respect to the metric $g_{\mu\nu}$ and matter fields $\chi$, as done in the appendix \ref{app-b} for the physical metric $\mathfrak{g}_{\mu\nu}=e^{2\alpha}g_{\mu\nu}$.

It has been shown in \cite{quiros-2025-b} that, under the above considerations, the gravitational action in \eqref{tot-action}, can be written in the following way:

\begin{align} S_\text{grav}=\frac{1}{2}\int d^4x\sqrt{-\hat{\mathfrak{g}}}\left[\hat{\mathfrak{R}}-(\hat{\mathfrak{D}}\hat\Phi)^2-2\hat{{\cal V}}\right],\label{eh-action-pact}\end{align} where the following conformal invariant scalar field,

\begin{align} \hat\Phi=\int\frac{\sqrt{\omega(\phi)+3/2}}{\phi}\,d\phi,\label{Phi}\end{align} has been introduced. In addition, it has also been shown that the following equivalence takes place \cite{quiros-2025-a} (see also \cite{quiros-2025-b}):

\begin{align} \sqrt{-g}\,L_m(\chi,\der\chi,g_{\mu\nu})\equiv\sqrt{-\hat{\mathfrak{g}}}\,L_m(\hat\Psi,\hat{\mathfrak{d}}\hat\Psi,\hat{\mathfrak{g}}_{\mu\nu})\;\Rightarrow\;S_m[\chi,\der\chi,g_{\mu\nu}]\equiv S_m[\hat\Psi,\hat{\mathfrak{d}}\hat\Psi,\hat{\mathfrak{g}}_{\mu\nu}].\label{mat-lag-equal}\end{align} Therefore, the total action \eqref{tot-action} is equivalent to the following action:

\begin{align} S_\text{tot}=&\frac{1}{2}\int d^4x\sqrt{-\hat{\mathfrak{g}}}\left[\hat{\mathfrak{R}}-(\hat{\mathfrak{D}}\hat\Phi)^2-2\hat{{\cal V}}\right]+\int d^4x\sqrt{-\hat{\mathfrak{g}}}\,L_m(\hat\Psi,\hat{\mathfrak{d}}\hat\Psi,\hat{\mathfrak{g}}_{\mu\nu}),\label{tot-lag'}\end{align} which is written in a covariant way in the configuration space. The total STG action in the parametrization \eqref{tot-lag'} is another expression of the passive approach to CTs for the theory \eqref{tot-action}. It is also not suitable for the investigation of the conformal symmetry.

Summarizing: The use of conformal form-invariant quantities to describe conformal form-invariant laws of gravity is not a well-posed problem. It is inevitably linked with the passive approach to CTs, which is useless for the study of the physical and phenomenological consequences of conformal symmetry.

%-------------------------------------

%%%%%%%%%%%%%%%%%%%%%%%%%%%%%%%%%%%%%%%%%%%%

\section{Discussion and concluding remarks}
\label{sect-discu}

%%%%%%%%%%%%%%%%%%%%%%%%%%%%%%%%%%%%%%%%%%%

Let us briefly summarize the main results of the present investigation and comment on them.

\begin{enumerate}

%---------item 1-----------

\item We have shown that the action of STG theory in the parametrization introduced in \cite{flanagan_cqg_2004}:

\begin{align} S=\frac{1}{2}\int d^4x\sqrt{-g}\left[AR-B(\der\phi)^2-2V\right]+\int d^4x\sqrt{-g}\,L_m\left(\chi,e^{2\alpha}g_{\mu\nu}\right),\nonumber\end{align} where $A=A(\phi)$, $B=B(\phi)$, and $V=V(\phi)$, is not conformal form-invariant in general. It is conformal form-invariant only for radiation. The matter action breaks the conformal symmetry even for massless fermions. To include any background matter fields, we must consider one of the following possibilities instead:

\begin{align} S=\frac{1}{2}\int d^4x\sqrt{-g}\left[AR-B(\der\phi)^2-2V\right]+\int d^4x\sqrt{-\mathfrak{g}}\,L_m\left(\Psi,\mathfrak{g}_{\mu\nu}\right),\label{c-1}\end{align} where $\mathfrak{g}_{\mu\nu}\equiv e^{2\alpha}g_{\mu\nu}$ and $\Psi\equiv e^{w_\chi\alpha}\chi$ are conformal invariant fields, or

\begin{align} S=\frac{1}{2}\int d^4x\sqrt{-g}\left[AR-B(\der\phi)^2-2V\right]+\int d^4x\sqrt{-g}\,L_m\left(\chi,g_{\mu\nu}\right).\label{c-2}\end{align} Both actions are conformal form-invariant under generalized CTs \eqref{gauge-t} and \eqref{flan-t}. 

In the form \eqref{c-1}, the STG action is suitable for the analysis under the passive approach to CT, while in the form \eqref{c-2}, it is suitable for the analysis under the active approach instead.

%-----------item 2------------

\item Following the PACT we have demonstrated that, in terms of the physical metric $\mathfrak{g}_{\mu\nu}$ and of the conformal invariant scalar $\vphi$ (see their definitions in equation \eqref{phys-g-vphi}), the action \eqref{c-1} can be written in JFBD parametrization:

\begin{align} S=\frac{1}{2}\int d^4x\sqrt{-\mathfrak{g}}\left[\vphi\mathfrak{R}-{\cal W}\frac{(\mathfrak{V}\vphi)^2}{\vphi}-2{\cal V}\right]+\int d^4x\sqrt{-\mathfrak{g}}\,L_m\left(\Psi,\mathfrak{g}_{\mu\nu}\right).\label{c-3}\end{align} Since in \eqref{c-3} the auxiliary fields, such as the metric $g_{\mu\nu}$, the scalar field $\phi$, and the matter fields $\chi$, do not appear explicitly, the conformal transformation \eqref{conf-t} plus the simultaneous reparametrizations \eqref{flan-t} amount to the identity transformation of the physical fields: $\mathfrak{g}_{\mu\nu}\rightarrow\mathfrak{g}_{\mu\nu}$, $\vphi\rightarrow\vphi$, ${\cal W}\rightarrow{\cal W}$, and $\Psi\rightarrow\Psi$. This means that conformal invariance is not an actual symmetry of \eqref{c-3}. Otherwise, the PACT is not suitable for the study of conformal symmetry.

%-------------item 3------------------

\item If we follow the AACT, the action \eqref{c-2} is the starting point of the analysis. We have shown that if we introduce a new parametrization: $\vphi=A(\phi)$ and $\omega(\vphi)=A(\phi)B(\phi)/A'^2(\phi)$ (here $A'\equiv A_{,\phi}$), this action is written in JFBD parametrization;

\begin{align} S_\text{tot}=\frac{1}{2}\int d^4x\sqrt{-g}\left[\vphi R-\frac{\omega}{\vphi}(\der\vphi)^2-2V(\vphi)\right]+\int d^4x\sqrt{-g}\,L_m(\chi,g_{\mu\nu}).\label{c-4}\end{align} The latter action and the derived EOMs are form-invariant under the generalized conformal transformation

\begin{align} g_{\mu\nu}\rightarrow\hat g_{\mu\nu}=\Omega^2g_{\mu\nu},\;\vphi\rightarrow\hat\vphi=\Omega^{-2}\vphi,\;\omega\rightarrow\hat\omega=\frac{\omega+6\frac{\Omega_{,\vphi}}{\Omega}\,\vphi\left(1-\frac{\Omega_{,\vphi}}{\Omega}\,\vphi\right)}{\left(1-2\frac{\Omega_{,\phi}}{\Omega}\,\phi\right)^2},\;m\to\hat m=\Omega^{-1}m,\label{c-5}\end{align} in the same way as \eqref{c-2} (and the derived EOMs \eqref{einst-eom} and \eqref{phi-eom}) is form-invariant under simultaneous CT \eqref{conf-t} and \eqref{flan-t}. The total actions \eqref{c-2} and \eqref{c-4} are fully equivalent.

For the vacuum case, invariance of \eqref{c-4} under generalized CT \eqref{c-5} was previously studied in \cite{faraoni_1998} (see also Appendix A of \cite{wands-rev-1999}). This means that the parametrization of \cite{flanagan_cqg_2004} is not more general than the usual JFBD parametrization if, in the latter case, consider form-invariance under the generalized conformal transformations \eqref{c-5}, which include a suitable transformation of the coupling parameter $\omega$. Notice that, since in STG theory $\omega=\omega(\vphi)$, any transformation of the BD scalar $\vphi$ induces a transformation of the coupling parameter. Hence, the natural way is to include a suitable transformation of $\omega$ as in \eqref{c-5}.

%--------------item 4-----------

\item An important, apparently trivial aspect discussed in the present document is counting degrees of freedom. Conformal symmetry implies an additional DOF: In addition to the four degrees of freedom to make coordinate transformations, there is a new DOF to choose either the BD scalar field $\phi=\phi(x)$ or one of the components of the metric. Here, for definiteness, we have chosen to fix the scalar field $\phi=\phi(x)$ to exhaust the additional DOF. As a matter of fact, the situation is not as simple as we just said. According to our previous analysis, the KG-type EOM obtained from the variation of the action \eqref{flan-lag'} with respect to the scalar field, taking into account the trace of the Einstein-type EOM \eqref{einst-eom}, can be written in the following way:

\begin{align} \nabla^2\phi+\beta(\phi)(\der\phi)^2=0,\label{phi-eom'}\end{align} where the parameter $\beta=\beta(\phi)\equiv d\ln\sqrt{2AB+3A'^2}/d\phi$. Hence, we have the freedom to choose the parameters $A=A(\phi)$ and $B=B(\phi)$, so that $\beta=\beta(\phi)$ is known. After this, it remains to solve the EOM \eqref{phi-eom'}. This amounts to fixing the scalar field $\phi$. Notice, however, that the obtained solution does not have dynamical implications for the gravitational interactions of matter. Hence, in practice, we have the freedom to choose any $\phi=\phi(x)$ we want. Every choice leads to a different global gravitational state, i.e., to a different phenomenology (see \cite{quiros-2025-b} for a much more detailed discussion on this subject). This is equivalent to the freedom to choose any pair $(A(\phi),B(\phi))$.

%--------------item 5-----------

\item Another interesting consequence of our study is that, according to the AACT, the fields that suffer the conformal transformation are themselves physically meaningful. In particular, each set of fields $\{g_{\mu\nu},\phi,\chi\}$ represents a gravitational state. The generalized conformal transformations \eqref{c-5} relate two different gravitational states

\begin{align} {\cal S}_g:(\chi,\phi,g_{\mu\nu})\Leftrightarrow\,\hat{\cal S}_{\hat g}:(\hat\chi,\hat\phi,\hat g_{\mu\nu}),\nonumber\end{align} which carry different phenomenological signatures, although both are described by the same gravitational laws.

\end{enumerate}

One of the main conclusions of our study is that the form in which the matter action is presented should be carefully investigated, and the fact that the action of matter fields in the standard form is already conformal form-invariant, must be taken into consideration. In this regard, there is not much freedom in the form of matter action that preserves conformal invariance. Only the following equivalent matter actions are conformal form-invariant:

\begin{align} S_m=\int d^4x\sqrt{-g}\,L_m\left(\chi,\der\chi,g_{\mu\nu}\right)\equiv\int d^4x\sqrt{-\mathfrak{g}}\,L_m\left(\Psi,\mathfrak{d}\Psi,\mathfrak{g}_{\mu\nu}\right),\nonumber\end{align} where $L_m\left(\chi,\der\chi,g_{\mu\nu}\right)$ is the standard Lagrangian of matter fields. One form is associated with the active approach to conformal transformations, while the other one is associated with the passive approach instead. Only within the framework of AACT conformal symmetry carries phenomenological consequences.

%---------------final comment---------------

Finally, let us comment on the possibility that conformal invariance may be an actual symmetry of the classical laws of gravity. If the standard model (SM) of particles and fields is modified in such a way as to encompass point-dependent masses \cite{nicolai-2007, bars-2014-a}, and additionally incorporates conformal symmetry in the variational process so that the correct EOM are derived \cite{quiros-2025-a}, then several of the arguments against conformal symmetry as an actual symmetry in physics are overcome.

%------------acknowledgments-----------------

{\bf Acknowledgments} The authors acknowledge FORDECYT-PRONACES-CONACYT for support of the present research under grant CF-MG-2558591.  

%-------------------------------------------

%%%%%%%%%%%%%%%%%%%%%%%%%%
%%%%%%%%%%%%%%%%%%%%%%%%%%

\appendix

%%%%%%%%%%%%%%%%%%%%%%%%%
%%%%%%%%%%%%%%%%%%%%%%%%%

%------Ward id------------------

%%%%%%%%%%%%%%%%%%%%%%%%%%%%%%%%%%%%%%%%%%%%

\section{Derivation of the Ward identity}
\label{app-a}

%%%%%%%%%%%%%%%%%%%%%%%%%%%%%%%%%%%%%%%%%%%%

The derivation of the Ward identity for the Lagrangian of matter fields in the case where the BD field $\phi$ has squared mass dimension can be found in \cite{quiros-2025-b}. Under the parametrization of \cite{flanagan_cqg_2004}, it is the parameter $A=A(\phi)$ that has squared mass dimension. Hence, one may guess that the mass parameter $m=\kappa\,\sqrt{A}$; that is, $m=m(A)$ and, consequently, the energy density of a perfect fluid, $\rho=\rho(A)$, both are explicit functions of the parameter $A$, so they are implicit functions of $\phi$. The Lagrangian density of matter in \eqref{flan-lag'}: ${\cal L}_m=\sqrt{-g}\,L_m\left(\chi,\der\chi,g_{\mu\nu},A\right)$ is conformal form-invariant.\footnote{The explicit dependence of the matter Lagrangian density on $A$ is clearly understood if we consider, for instance, the Dirac fermion Lagrangian density: ${\cal L}_\text{dirac}=\sqrt{-g}\bar\psi\left(i\cancel{\cal D}+m\right)\psi,$ where $\psi$ is the Dirac fermion spinor ($\bar\psi$ is the adjoint spinor), and $\cancel{\cal D}:=\gamma^\mu{\cal D}_\mu$. The mass of the fermion $m=m(A)$, introduces the explicit dependence on the parameter $A$. In the case of the Lagrangian density of a perfect fluid, the explicit dependence on $A$ is introduced by the energy density, or by the pressure of the perfect fluid.} Therefore, under the following infinitesimal conformal transformations:

\begin{align} \delta g_{\mu\nu}=2\theta\,g_{\mu\nu}\,\left(\delta g^{\mu\nu}=-2\theta\,g^{\mu\nu},\;\delta\sqrt{-g}=4\theta\sqrt{-g}\right),\;\delta A=-2\theta A,\label{inf-gauge-t}\end{align} where for convenience (and only temporarily) we have rescaled the conformal factor: $\Omega=e^\theta$, the overall variation of the Lagrangian density of matter satisfies $\delta{\cal L}_m=0$ (or, in general, $\delta{\cal L}_m=\nabla_\mu V^\mu$, where $V^\mu$ is some vector). We have, $\delta S_m=\int d^4x\,\delta{\cal L}_m=0$, so that under \eqref{inf-gauge-t}, $S_m\rightarrow S_m+\delta S_m=S_m$; that is, the matter action is invariant under \eqref{inf-gauge-t}. For the overall variation of the Lagrangian density of matter, we have\footnote{In classical gravitational problems, it is understood that the matter fields are apriori given; that is, the EOMs of the matter fields: $$\frac{\delta{\cal L}_m}{\delta\chi_i}=0\;\Rightarrow\;\frac{\der{\cal L}_m}{\der\chi_i}-\nabla_\mu\left[\frac{\der{\cal L}_m}{\der(\der_\mu\chi_i)}\right]=0,$$ are automatically satisfied.}

\begin{align} \delta{\cal L}_m=\frac{\delta{\cal L}_m}{\delta g^{\mu\nu}}\delta g^{\mu\nu}+\frac{\delta{\cal L}_m}{\delta A}\delta A=-2\theta\left(g^{\mu\nu}\frac{\delta{\cal L}_m}{\delta g^{\mu\nu}}+A\frac{\delta{\cal L}_m}{\delta A}\right)=0\;\Rightarrow\;\frac{\delta{\cal L}_m}{\delta\phi}=-\frac{A'}{A}g^{\mu\nu}\frac{\delta{\cal L}_m}{\delta g^{\mu\nu}},\label{var-lmat}\end{align} where, in the last step we took into account that $\delta A=A'\delta\phi$. Since, $\delta{\cal L}_m/\delta g^{\mu\nu}=-\sqrt{-g}\,T^{(m)}_{\mu\nu}/2$, i.e., $g^{\mu\nu}\delta{\cal L}_m/\delta g^{\mu\nu}=-\sqrt{-g}\,T^{(m)}/2$, then, finally, the Ward identity can be written in the following way:

\begin{align} \frac{\delta{\cal L}_m}{\delta\phi}=\frac{A'}{2A}\,\sqrt{-g}\,T^{(m)}.\label{app-a-ward-id}\end{align}

%---------basic elements--------------

%%%%%%%%%%%%%%%%%%%%%%%%%%%%%%%%%%%%%%%%%%%%%%%%%%%

\section{Relationships between the conformal invariant and the auxiliary fields}
\label{app-b}

%%%%%%%%%%%%%%%%%%%%%%%%%%%%%%%%%%%%%%%%%%%%%%%%%%%

In \eqref{phys-g} we have defined a conformal invariant metric tensor. Although this is not the only way in which a conformal invariant metric can be defined (for instance, the product $A(\phi)g_{\mu\nu}$ is conformal invariant as well) from the action of matter fields \eqref{flan-mat-action} it is evident that the matter fields follow geodesics of the conformal invariant metric $\mathfrak{g}_{\mu\nu}=e^{2\alpha}g_{\mu\nu}$, which we call the physical metric. Meanwhile, the metric tensor $g_{\mu\nu}$ is called an auxiliary metric.

We define a physical Riemann space $V_4:({\cal M}_4,\mathfrak{g}_{\mu\nu})$, where ${\cal M}_4$ is a four-dimensional manifold with the required properties. The affine connection of the physical $V_4$ coincides with the physical Levi-Civita connection (same as Christoffel symbols):

\begin{align} \mathfrak{G}^\lambda_{\mu\nu}=\frac{1}{2}\mathfrak{g}^{\lambda\kappa}\left(\der_\nu\mathfrak{g}_{\mu\kappa}+\der_\mu\mathfrak{g}_{\nu\kappa}-\der_\kappa\mathfrak{g}_{\mu\nu}\right).\label{phys-lc-c}\end{align} If insert the definition \eqref{phys-g} of the physical metric in the above equation one obtains the relationship between the physical LC connection and the Christoffel symbols of the auxiliary metric $\left\{^\lambda_{\mu\nu}\right\}$,

\begin{align} \mathfrak{G}^\lambda_{\mu\nu}=\left\{^\lambda_{\mu\nu}\right\}+L^\lambda_{\mu\nu},\label{aff-c-rel}\end{align} where

\begin{align} L^\lambda_{\mu\nu}\equiv\delta^\lambda_\mu\der_\nu\alpha+\delta^\lambda_\nu\der_\mu\alpha-g_{\mu\nu}\der^\lambda\alpha.\label{l-def}\end{align}

The physical Riemann curvature tensor, the corresponding Ricci tensor, and the curvature scalar are as follows.

\begin{align} \mathfrak{R}^\alpha_{\;\;\mu\beta\nu}&=R^\alpha_{\;\;\mu\beta\nu}+\nabla_\beta L^\alpha_{\;\;\mu\nu}-\nabla_\nu L^\alpha_{\;\;\beta\mu}+L^\alpha_{\;\;\beta\lambda}L^\lambda_{\;\;\nu\mu}-L^\alpha_{\;\;\nu\lambda}L^\lambda_{\;\;\beta\mu},\nonumber\\
\mathfrak{R}_{\mu\nu}&=R_{\mu\nu}+2\der_\mu\alpha\der_\nu\alpha-2g_{\mu\nu}(\der\alpha)^2-2\nabla_\mu\nabla_\nu\alpha-g_{\mu\nu}\nabla^2\alpha,\nonumber\\
\mathfrak{R}&=e^{-2\alpha}\left[R-6(\der\alpha)^2-6\nabla^2\alpha\right],\label{curv-q}\end{align} respectively. Alternatively,

\begin{align} R_{\mu\nu}&=\mathfrak{R}_{\mu\nu}+2\mathfrak{V}_\mu\alpha\mathfrak{V}_\nu\alpha-2\mathfrak{g}_{\mu\nu}(\mathfrak{V}\alpha)^2+2\left(\mathfrak{V}_\mu\mathfrak{V}_\nu+\frac{1}{2}\mathfrak{g}_{\mu\nu}\mathfrak{V}^2\right)\alpha,\nonumber\\
R&=e^{2\alpha}\left[\mathfrak{R}-6(\mathfrak{V}\alpha)^2+6\mathfrak{V}^2\alpha\right],\label{curv-q'}\end{align} where the covariant derivative operator $\mathfrak{V}_\mu$ in the physical Riemann space is defined in terms of the physical LC connection \eqref{phys-lc-c} and we have used the notation: $(\mathfrak{V}\alpha)^2\equiv\mathfrak{g}^{\mu\nu}\mathfrak{V}_\mu\alpha\mathfrak{V}_\nu\alpha$ and $\mathfrak{V}^2\equiv\mathfrak{g}^{\mu\nu}\mathfrak{V}_\mu\mathfrak{V}_\nu.$

The relationship between the tangent Minkowski space metric $\eta_{ab}$ ($a,b,c...=0,1,2,3$ are the tangent space or flat indices) and the physical metric is given by

\begin{align} \mathfrak{g}_{\mu\nu}=\eta_{ab}\mathfrak{e}^a_\mu\mathfrak{e}^b_\nu,\;\;\eta_{ab}=\mathfrak{g}_{\mu\nu}\mathfrak{e}^\mu_a\mathfrak{e}^\nu_b,\label{metric-rel}\end{align} where $\mathfrak{e}^a_\mu=e^\alpha\,e^a_\mu$ are the physical tetrad fields ($\mathfrak{e}^\mu_a=e^{-\alpha}\,e^\mu_a$). In a similar way the physical Dirac gamma matrices $\Gamma^\mu=e^{-\alpha}\gamma^\mu$ can be defined. We have that,

\begin{align} \gamma^a=\Gamma^\mu\mathfrak{e}^a_\mu.\label{d-gamma}\end{align} 

If we write the gauge covariant derivative ${\cal D}_\mu$ in terms of physical quantities, we obtain the following.

\begin{align} {\cal D}_\mu=\mathfrak{D}_\mu-\frac{3}{2}\,\der_\mu\alpha,\label{rel-der}\end{align} where we have defined the physical gauge covariant derivative,

\begin{align} \mathfrak{D}_\mu:=D_\mu-\frac{1}{2}\sigma_{ab}\mathfrak{e}^{b\nu}\mathfrak{V}_\mu\mathfrak{e}^a_\nu.\label{gauge-cov-der}\end{align} In the above equations, we use the standard definition of the gauge derivative $SU(2)\times U(1)$, $D_\mu$. Let us consider the gauge covariant derivative of the Dirac fermion field $\psi=e^{3\alpha/2}\Psi$, where $\Psi$ is a conformal invariant spinor:

\begin{align} \cancel{\cal D}\psi=\gamma^\mu{\cal D}_\mu\left(e^{3\alpha/2}\Psi\right)=e^\alpha\Gamma^\mu{\cal D}_\mu\left(e^{3\alpha/2}\Psi\right)=e^{5\alpha/2}\Gamma^\mu\mathfrak{D}_\mu\Psi=e^{5\alpha/2}\cancel{\mathfrak{D}}\Psi,\label{phys-g-der}\end{align} and we took into account that

\begin{align} \left(\mathfrak{D}_\mu-\frac{3}{2}\,\der_\mu\alpha\right)e^{3\alpha/2}\Psi=e^{3\alpha/2}\mathfrak{D}_\mu\Psi.\nonumber\end{align}

We want to underline that, under the simultaneous conformal transformation \eqref{conf-t} and $\psi\rightarrow\Omega^{-3/2}\psi$, ${\cal D}_\mu\psi$ transforms as: ${\cal D}_\mu\psi\rightarrow\Omega^{-3/2}{\cal D}_\mu\psi$, while

\begin{align} \mathfrak{D}_\mu\psi\rightarrow\Omega^{-3/2}\left(\mathfrak{D}_\mu-\frac{3}{2}\der_\mu\ln\Omega\right)\psi.\label{app-b-eq}\end{align}

%----------conf inv matter---------------

%%%%%%%%%%%%%%%%%%%%%%%%%%%%%%%%%%%%%%%%%%%%%%%%

\section{Conformal invariant matter fields}
\label{app-c}

%%%%%%%%%%%%%%%%%%%%%%%%%%%%%%%%%%%%%%%%%%%%%%%

If the conformal invariant fermion field $\Psi=e^{-3\alpha/2}\psi$ is coupled to the conformal invariant metric $\mathfrak{g}_{\mu\nu}$ given by \eqref{phys-g}, we obtain the following equality:

\begin{align} \sqrt{-g}\,\bar\psi\cancel{\cal D}\psi=\sqrt{-\mathfrak{g}}\,\bar\Psi\cancel{\mathfrak{D}}\Psi,\label{equal}\end{align} where $\cancel{\mathfrak{D}}=\Gamma^\mu\mathfrak{D}_\mu$, $\Gamma^\mu=e^{-\alpha}\gamma^\mu$ are the physical Dirac gamma matrices, $\mathfrak{e}^a_\mu=e^\alpha\,e^a_\mu$ the corresponding tetrad, $\mathfrak{D}_\mu$ is the physical gauge covariant derivative, $\mathfrak{V}_\mu$ is the covariant derivative of the physical metric, etc.\footnote{For notation and details of the demonstration of equality \eqref{equal}, see Appendix \ref{app-b}.} While the left-hand side (LHS) of equality \eqref{equal} is manifestly conformal invariant, its RHS is trivially conformal invariant since under \eqref{conf-t}; $\sqrt{-\mathfrak{g}}\rightarrow\sqrt{-\mathfrak{g}}$, $\Psi\rightarrow\Psi$ ($\bar\Psi\rightarrow\bar\Psi$) and $\cancel{\mathfrak{D}}\rightarrow\cancel{\mathfrak{D}}$. In this case, the conformal transformation is a fictitious symmetry since it affects only the auxiliary fields $g_{\mu\nu}$ and $\psi$, but not the physical fields $\mathfrak{g}_{\mu\nu}$ and $\Psi$.

Next, we investigate the mass term,

\begin{align} \sqrt{-g}\,m\bar\psi\psi,\label{m-term}\end{align} where, under the conformal transformation, the mass parameter transforms like in \eqref{mass-t}. If we substitute the following relationships: $\psi=e^{3\alpha/2}\Psi$ and $\sqrt{-g}=e^{-4\alpha}\sqrt{-\mathfrak{g}}$, back into \eqref{m-term}, we obtain the following equality:

\begin{align} \sqrt{-g}\,m\bar\psi\psi=\sqrt{-\mathfrak{g}}\,\mathfrak{m}\bar\Psi\Psi,\label{equ-1}\end{align} where in the RHS, only conformal invariant quantities are involved. This includes the conformal invariant mass parameter $\mathfrak{m}=e^{-\alpha}m$. Therefore, the action of the Dirac fermion can be written in terms of conformal invariant physical quantities in the following way:

\begin{align} S_f=\int d^4x\sqrt{-\mathfrak{g}}\,\bar\Psi\left(i\cancel{\mathfrak{D}}+\mathfrak{m}\right)\Psi.\label{cinv-dirac-action}\end{align} Written in this way the action of physical matter fields is not actually conformal form-invariant because, as far as the auxiliary fields are removed from the covariant description, conformal form-invariance is a fictitious symmetry.

%------------conformal invariant fields--------------

Consider that the Proca field is coupled to the conformal invariant metric $\mathfrak{g}_{\mu\nu}=e^{2\alpha}g_{\mu\nu}$. Let us introduce the following conformal invariant quantities:

\begin{align} \mathfrak{F}_{\mu\nu}=F_{\mu\nu},\;\mathfrak{F}^{\mu\nu}=e^{-4\alpha}F^{\mu\nu},\;{\cal A}_\mu=A_\mu,\,{\cal A}^\mu=e^{-2\alpha}A^\mu,\,\mathfrak{m}_p=e^{-\alpha}m_p.\label{def-proca}\end{align} The Proca action can be written as follows:

\begin{align} S_p=\int d^4x\sqrt{-\mathfrak{g}}\,\left(\frac{1}{4}\mathfrak{F}^2+\frac{1}{2}\mathfrak{m}^2_p{\cal A}^2\right).\label{cinv-proca-action}\end{align} In this case, conformal form-invariance is a fictitious symmetry since CT \eqref{conf-t} becomes the identity transformation:

\begin{align} \sqrt{-\mathfrak{g}}&\rightarrow\sqrt{-\mathfrak{g}},\,\mathfrak{F}^2\rightarrow\mathfrak{F}^2,\;\mathfrak{m}^2_p\rightarrow\mathfrak{m}^2_p,\;{\cal A}^2\rightarrow{\cal A}^2.\nonumber\end{align} 

%-------------------------------

Now we shall check that the equality \eqref{flan-lm-eq} is wrong, in general.\footnote{The same is valid for \eqref{jarv-lm-eq}.} Let us write the Proca action in the form \eqref{flan-mat-action} proposed in \cite{flanagan_cqg_2004}. For the Proca field, we have $S_m\left[\chi,\mathfrak{g}_{\mu\nu}\right]$, where the matter field is $\chi=A_\mu$. Hence, we must replace $g_{\mu\nu}\rightarrow\mathfrak{g}_{\mu\nu}$ while keeping the vector field $A_\mu$ and its mass $m_p$ unchanged:

\begin{align} S_p\left[A_\mu,\mathfrak{g}_{\mu\nu}\right]=\int d^4x\sqrt{-\mathfrak{g}}\,&\left(\frac{1}{4}\mathfrak{g}^{\mu\lambda}\mathfrak{g}^{\nu\kappa}F_{\mu\nu}F_{\lambda\kappa}+\frac{1}{2}m^2_p\mathfrak{g}^{\mu\nu}A_\mu A_\nu\right).\nonumber\end{align} Taking into account the definitions \eqref{def-proca}, the above action is written in the following compact form:

\begin{align} S_p\left[A_\mu,\mathfrak{g}_{\mu\nu}\right]=\int d^4x\sqrt{-\mathfrak{g}}\,\left(\frac{1}{4}\mathfrak{F}^2+\frac{1}{2}m^2_p{\cal A}^2\right).\label{mix-proca-action}\end{align} Notice the subtle difference between \eqref{cinv-proca-action} and \eqref{mix-proca-action}: the mass term breaks the trivial conformal form-invariance of \eqref{mix-proca-action} since, under the conformal transformation $m_p\rightarrow\Omega^{-1}m_p$. For EM radiation, which corresponds to the limit $m_p\rightarrow 0$ in the above equations, the trivial conformal form-invariance of the matter action is a correct result.

The demonstration that the matter action in the form considered in \cite{flanagan_cqg_2004}; $S_m=S_m[\chi,\mathfrak{g}_{\mu\nu}]$, is not conformal invariant for the action of a Dirac fermion, is even simpler. In this case, in the Dirac action we must replace $g_{\mu\nu}\rightarrow\mathfrak{g}_{\mu\nu}$, $e^a_\mu\rightarrow\mathfrak{e}^a_\mu$, etc., while $\chi=\psi$. Since the fermion spinor $\psi$ has a nonvanishing conformal weight, the equality \eqref{flan-lm-eq} is not satisfied. Nevertheless, let us temporarily assume that the Dirac action in the form \eqref{flan-mat-action} is correct. We have

\begin{align} S_f=\int d^4x\sqrt{-\mathfrak{g}}\,\bar\psi\left(i\cancel{\mathfrak{D}}+m\right)\psi,\label{dirac-action'}\end{align} where the physical gauge covariant derivative $\mathfrak{D}_\mu$ is defined in terms of the physical metric $\mathfrak{g}_{\mu\nu}$ and of the physical tetrad $\mathfrak{e}^a_\mu$ (see equation \eqref{gauge-cov-der} of Appendix \ref{app-b}). According to \eqref{app-b-eq}, $\bar\psi\cancel{\mathfrak{D}}\psi\rightarrow$ $\Omega^{-3}\bar\psi(\cancel{\mathfrak{D}}-3\Gamma^\mu\der_\mu\ln\Omega)\psi$ so that the action \eqref{dirac-action'} is not conformal form-invariant. This is confirmed by the mass term since, under \eqref{conf-t}, $\sqrt{-\mathfrak{g}}\,m\bar\psi\psi\rightarrow\Omega^{-4}\sqrt{-\mathfrak{g}}\,m\bar\psi\psi$. 

We see that if we follow the particular form of the matter action assumed in \cite{flanagan_cqg_2004, jarv_2015}, even for massless fermions, the matter action is not conformal form-invariant. The same holds true if we consider either fermion or radiation fields with nonvanishing mass. The results of the present discussion are also valid for equality \eqref{kalo}.

It must be underlined that it is required by consistency of a given metric theoretical framework that the action of given STG theory be written in terms of only that metric tensor to which the matter fields are minimally coupled. This is the metric that defines the proper time measured by actual physical clocks. In addition, point particles and matter fields follow the Riemannian geodesics of precisely this metric tensor. Why then to look for a metric tensor different from the one with macroscopic measurable significance to account for the shape of gravity?

%-------------------------------

%%%%%%%%%%%%%%%%%%%%%%%%%%%%%

%%%%%%%%%%%%%%%%%%%%%%%%%%

%%%%%%%%%%%%%%%%%
%%%%%%%%%%%%%%%%%

\end{document}